\documentclass[sigconf,authorversion,nonacm]{acmart} 

\AtBeginDocument{%
  }

\usepackage{subfigure}
\usepackage{stfloats}
\usepackage{bbm} 
\usepackage{multirow}
\usepackage{array}
\usepackage{tabularray}
\renewcommand\arraystretch{1.1}
\usepackage{booktabs}
\usepackage{makecell}
\usepackage{tabularx}

\begin{document}

\title{Think before You Leap: Content-Aware Low-Cost Edge-Assisted Video Semantic Segmentation}

%
\author{Mingxuan Yan}
\email{mingxuanyan@hust.edu.cn}
\affiliation{%
 \institution{Huazhong University of Science and Technology}
 \city{}
 \country{}
}

\author{Yi Wang}
\email{yi_wang_@hust.edu.cn}
\affiliation{%
 \institution{Huazhong University of Science and Technology}
 \city{}
 \country{}
}

\author{Xuedou Xiao}
\email{xuedouxiao@hust.edu.cn}
\affiliation{%
 \institution{Huazhong University of Science and Technology}
 \city{}
 \country{}
}

\author{Zhiqing Luo}
\email{zhiqing_luo@hust.edu.cn}
\affiliation{%
 \institution{Huazhong University of Science and Technology}
 \city{}
 \country{}
}

\author{Jianhua He}
\email{j.he@essex.ac.uk}
\affiliation{%
 \institution{University of Essex}
 \city{}
 \country{}
}

\author{Wei Wang}
\authornote{Corresponding author.}
\email{weiwangw@hust.edu.cn}
\affiliation{%
 \institution{Huazhong University of Science and Technology}
 \city{}
 \country{}
}

\renewcommand{\shortauthors}{Mingxuan Yan et al.}

\newcommand{\sysname}{{\textit{Penance}}}
\newcommand{\iotdevice}{{Raspberry Pi 4B}}
\begin{abstract}
Offloading computing to edge servers is a promising solution to support growing video understanding applications at resource-constrained IoT devices. Recent efforts have been made to enhance the scalability of such systems by reducing inference costs on edge servers. However, existing research is not directly applicable to pixel-level vision tasks such as video semantic segmentation (VSS), partly due to the fluctuating VSS accuracy and segment bitrate caused by the dynamic video content. In response, we present \textit{Penance}, a new edge inference cost reduction framework. By exploiting softmax outputs of VSS models and the prediction mechanism of H.264/AVC codecs, \textit{Penance} optimizes model selection and compression settings to minimize the inference cost while meeting the required accuracy within the available bandwidth constraints. We implement \textit{Penance} in a commercial IoT device with only CPUs. Experimental results show that \textit{Penance} consumes a negligible 6.8\% more computation resources than the optimal strategy while satisfying accuracy and bandwidth constraints with a low failure rate.

\end{abstract}


%

\maketitle


\section{Introduction}
\label{Introduction}

Recent years have witnessed a growing demand for IoT video analytics. The global IoT video analytics market is projected to increase from \$5.32 billion in 2021 to \$28.37 billion by 2029 \cite{noauthor_video_nodate}. Pixel-level video labeling task such as video semantic segmentation (VSS) is at the heart of IoT video analytics applications ranging from drones \cite{bhatnagar_drone_2020, chakravarthy_dronesegnet_2022}, video surveillance \cite{guerrero_tello_convolutional_2022, muhadi2021deep, pierard_mixture_2023}, augmented-reality \cite{anil_chandra_naidu_matcha_2021_nodate}, traffic scene understanding \cite{feng_deep_2021, Can_2021_ICCV}, to traffic safety \cite{chen_importance-aware_2019, Baheti_2020_CVPR_Workshops}. 
However, it is noted that the widespread adoption of VSS will heavily rely on deep learning techniques \cite{mo_review_2022} and that the increasing computational complexity of deep neural networks (DNNs) poses a significant challenge for video understanding on resource-constrained IoT devices.

An emerging solution is to offload the video analytics task to edge/cloud servers. In typical edge-assisted VSS systems, IoT devices stream encoded video segments to the server through a network link with limited bandwidth. The server then performs VSS to meet the users' accuracy requirements. However, there are increasing concerns over the scalability of such systems. The  state-of-the-art semantic segmentation models \cite{zhao_icnet_2018,zhao_pyramid_2017, yu_bisenet_2018, wang_swiftnet_2021} are computationally expensive, even for edge/cloud servers \cite{yuan_infi_2022, li_reducto_2020, jiang_joint_2021}. Furthermore, one server typically serves multiple users \cite{li_reducto_2020, wang_joint_2020}, which amplifies the computational overhead. Although efforts  \cite{li_reducto_2020, chen_glimpse_2015, yuan_infi_2022, jiang_chameleon_2018} have been made to reduce the edge inference cost by reusing cached predictions, these methods cannot be applied to pixel-level vision tasks as the pixel association varies greatly between frames, leading to drastic degradation of accuracy \cite{zhang_edge_2022}.

A promising solution to mitigate edge inference costs is to switch between multiple vision models with different costs. Specifically, this approach selects the best combination of the vision model version and compression settings to ensure minimal inference cost while satisfying accuracy and bandwidth constraints. Despite its success in image classification and object detection \cite{wang_joint_2020, jiang_joint_2021, zhang_adaptive_2022}, exploiting this approach for edge-assisted VSS faces grand challenges due to the following major issues.

\textit{(\romannumeral1) How to manage the fluctuating VSS performance caused by dynamic video content?} 
To switch between different models, the priority is to monitor their runtime accuracy. However, our measurements indicate that \textit{VSS models experience large accuracy fluctuations over short time windows of tens of seconds due to the changes of video contents}.
Recent works \cite{wang_joint_2020, zhang_adaptive_2022, li_reducto_2020} adapt the accuracy function by exploiting cheap features such as object sizes and edges, but they cannot be extended to VSS as semantic segmentation does not have explicit and concentrated regions of interest (RoIs) \cite{xiao_dnn-driven_2022}. Other works \cite{jiang_chameleon_2018,li_reducto_2020, zhang_awstream_2018} reprofile the edge vision model periodically (tens of minutes) and require raw frames fed to the edge server. Increasing reprofile frequency will overwhelm the limited network capacity and is computationally expensive. 

\textit{(\romannumeral2) How to tune the codec compression settings?} 
Before being transmitted over a limited network link, video segments must be compressed with codecs such as H.264/AVC by adjusting compression settings, including frame resolution and quantization parameter (QP). Previous works \cite{wang_joint_2020, zhang_adaptive_2022} map compression settings to segment bitrate by assuming a fixed relationship between them, \textit{ignoring the variation of video contents that can significantly affect the segment bitrate} as concluded in our measurements. Simply iterating and encoding all possible compression settings would be computationally infeasible. Though recent work \cite{murad_dao_2022} employs constant bitrate (CBR) to compress video segments using given bitrates to bypass this problem, it overlooks the trade-off between resolution and QP in bitrate saving and cost reduction, which offers more space for optimization.

\begin{figure*}[t]
    \setlength{\abovecaptionskip}{0cm}
	\setlength{\belowcaptionskip}{-0.2cm}
	\begin{minipage}{0.49\linewidth}
		\centering
		\includegraphics[width=0.9\linewidth]{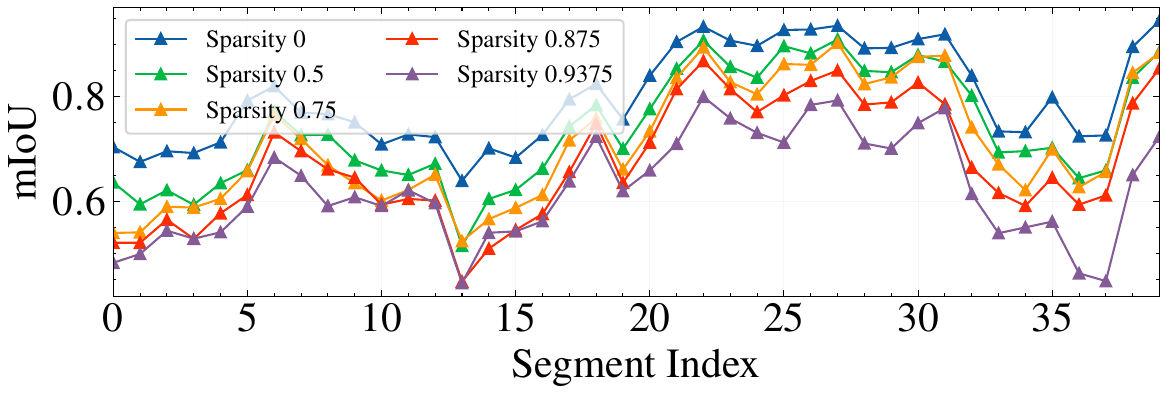}
		\caption{Showcase of the VSS accuracy fluctuation}
		\label{fig-acc-var-showcase}
	\end{minipage}\enspace	
	\begin{minipage}{0.49\linewidth}
		\centering
		\includegraphics[width=0.9\linewidth]{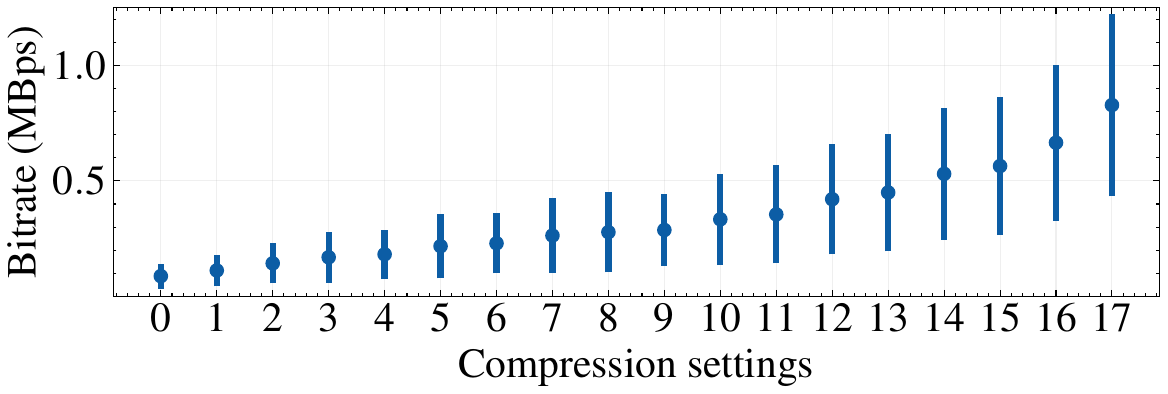}
		\caption{Bandwidth usage distributions}
		\label{fig-bandwidth-var}
	\end{minipage}\enspace	
    \vspace{-0.2cm}
\end{figure*}

\begin{figure*}[t]
    \setlength{\abovecaptionskip}{0cm}
	\setlength{\belowcaptionskip}{-0.4cm}
	\begin{minipage}{0.32\linewidth}
			\centering
			\includegraphics[width=0.9\linewidth]{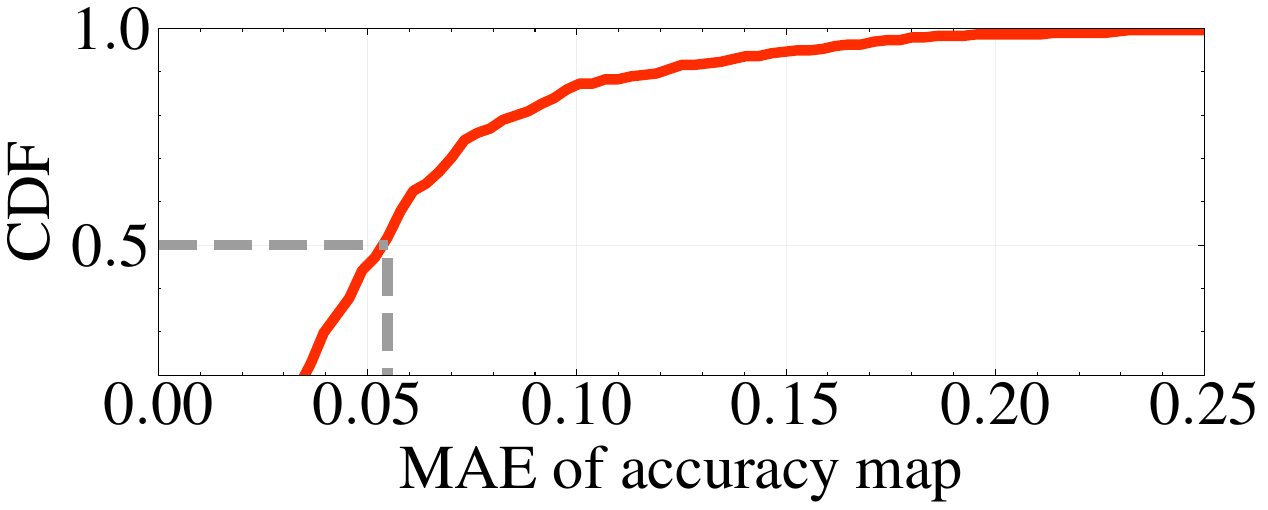}
			\caption{MAE of the reprofiled accuracy functions in the following 20 seconds}
			\label{fig-mae-accmap}
	\end{minipage}\enspace	
	\begin{minipage}{0.32\linewidth}
			\centering
			\includegraphics[width=0.9\linewidth]{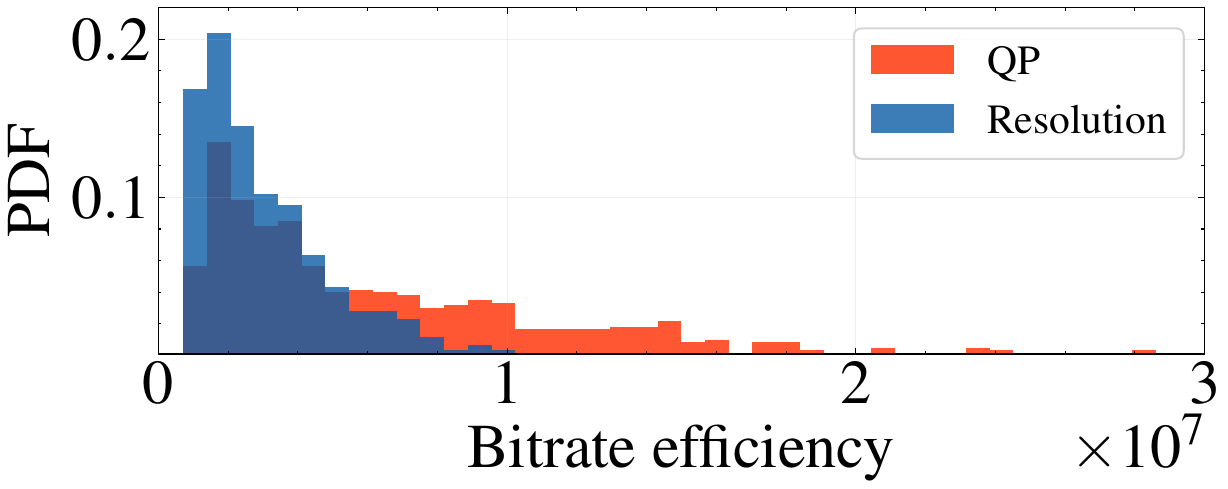}
			\caption{Compare bitrate efficiencies of QP and resolution}
			\label{fig-comp-effi}
	\end{minipage}\enspace	
	\begin{minipage}{0.32\linewidth}
		\centering
		\includegraphics[width=0.9\linewidth]{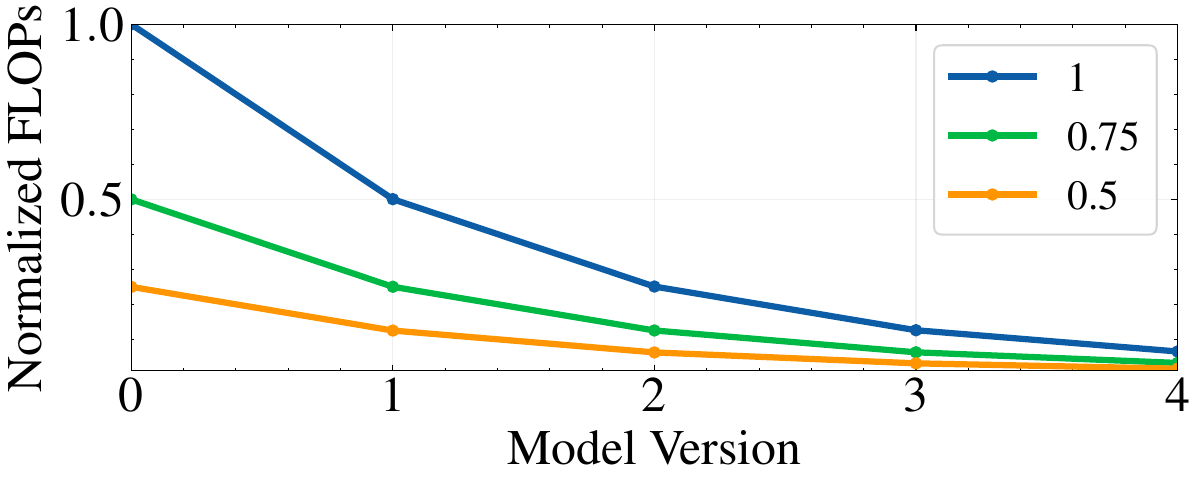}
		\caption{FLOPs of different resolution scaling factors}
		\label{fig-flops-rs}
	\end{minipage}\enspace	
\end{figure*}

\textit{(\romannumeral3) How to configure video segments?} 
Optimizing the configuration of the VSS model version and compression settings is challenging. Previous works \cite{wang_joint_2020, jiang_joint_2021, zhang_adaptive_2022} formulate and solve the cost reduction as a mixed-integer programming problem requiring accurate configuration mapping to model accuracies. However, the complex, \textit{context-dependent accuracy function of VSS makes accurate modeling challenging}, as mentioned above. Moreover, the optimization is even more challenging due to the \textit{inherent conflict between optimization goals} of minimizing the inference cost and satisfying the accuracy and bandwidth usage constraints.

Given the above practical issues, we present \sysname{}, the first content-aware and low-cost edge-assisted VSS system.
First, to monitor runtime VSS performance, \sysname{} exploits the most recent video frame's predicted softmax probabilities, from where a deep neural network (DNN) extracts an embedding that represents the runtime edge model performance. Second, to estimate the fluctuating segment bitrate, another special DNN is devised by exploiting the prediction mechanism of the H.264/AVC codecs to predict bandwidth usage for each compression setting using raw video frames. Finally, a deep reinforcement learning (DRL) model is presented to optimize the configuration for each segment, considering the accuracy and bandwidth constraints. \sysname{} is designed to be lightweight and can be deployed on general IoT devices equipped with only CPUs.

\textbf{Contributions.} 
\textit{(\romannumeral1)} 
We investigate the challenge and impact of dynamic video content on edge-assisted VSS systems.
\textit{(\romannumeral2)} 
We propose \sysname{}, the first low-cost edge-assisted VSS system for resource-constraint IoT devices by adapting both compression settings and edge model selection.
\textit{(\romannumeral3)} 
We implement \sysname{} on a commercial IoT device with only CPUs and evaluate its performance with baseline methods. The experiments show that our solution significantly lowers the edge inference cost while strictly adhering to all constraints.

The remainder of this paper is organized as follows. \S\ref{finding} introduces our measurement studies that motivates the design of \sysname{}. 
\S\ref{design} introduces the key ideas on the system design. 
\S\ref{eval} presents the implementation and evaluation of \sysname{}. \S\ref{related} gives a literature review of related works, followed by a conclusion in \S\ref{conclusion}.

\section{Measurement and Motivation}\label{finding}
\subsection{Measurement Setup}
\subsubsection{Semantic Segmentation Networks}
This paper adopts the popular PSPNet \cite{zhao_pyramid_2017} with the ResNet-50 backbone as the original VSS model. To obtain edge models with different inference costs, we resort to the DNN pruning technique \cite{ditschuneit_auto-compressing_2022} and derive 4 models from the original model with sparsity factors of 0.5, 0.75, 0.875, and 0.9375 respectively, whose computational overhead is inversely proportional to their sparsity. For simplicity, we name them as model 0 to 4, respectively.

\subsubsection{Dataset}\label{finding-dataset}
We conduct our measurement on BDD100k \cite{bdd100k}, a state-of-the-art large-scale diverse video dataset that contains 100K 1280x720 challenging videos with 40 seconds each, covering different scenarios such as urban and rural areas, highways, tunnels, etc. We first train and prune the 5 models with different sparsities on the training set. Then we randomly choose 20 videos spanning 800 seconds from the test set to conduct our measurement.
We split the videos into 1-second segments with a frame rate of 15 fps and compress them using standard H.264/AVC to obtain varying bandwidth usages. We choose quantization parameter (QP) and frame resolution as knobs to control video quality, which are typical video encoding settings in video analytics \cite{zhang_casva_2022, xiao_dnn-driven_2022, du_server-driven_2020}. Specifically, we compress the original videos into 18 variances, with QPs ranging from 20 to 30 and resolution scaling factors of 1, 0.75, and 0.5. We call the combinations of edge model versions and compression settings ``\textit{configurations}''.
As there is only one frame annotated in each video, we follow the practices of previous works \cite{zhang_awstream_2018,jiang_chameleon_2018} and regard predictions of the most expensive configuration (1280x720, QP=20 and model 0) as ground truths. We use the standard mean intersection over union (mIoU) metric to evaluate the performance of VSS models.

\begin{figure}[t]
    \setlength{\abovecaptionskip}{0.2cm}
	\setlength{\belowcaptionskip}{-0.6cm}
	\centering
	\includegraphics[width=\linewidth]{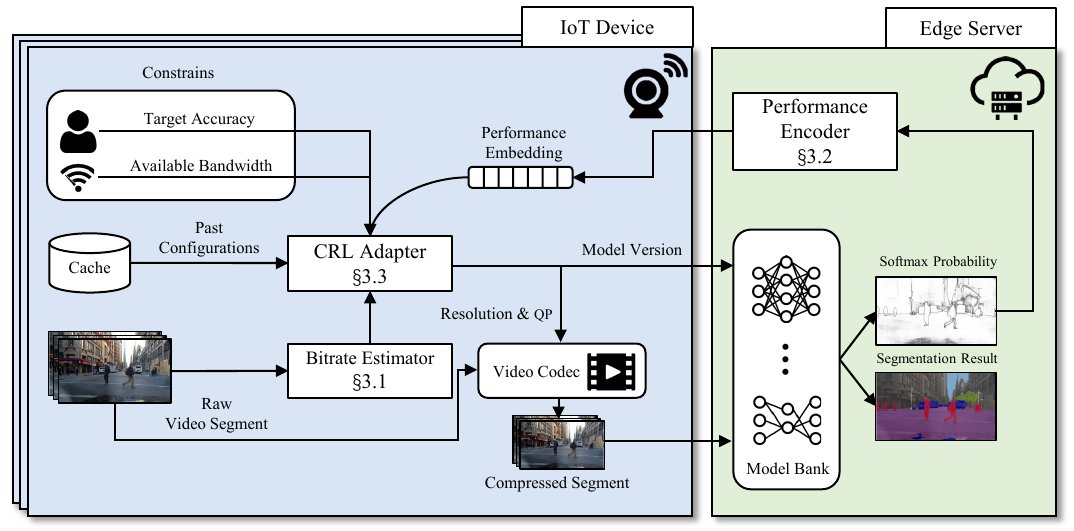}
	\caption{\sysname{} Overview}
	\Description{}
	\label{fig-se-illu-avc}
        \vspace{-0.1cm}
\end{figure}

\subsection{Impact of Dynamic Video Contents}
\subsubsection{Impact on Model Performances}
We first investigate the runtime performances of the edge VSS models. In Fig.~\ref{fig-acc-var-showcase}, we showcase the accuracy fluctuation of one 40-second video. It can be seen that the runtime model accuracy experiences drastic changes with time. We can further observe that the relative ratio between the accuracies of different models also changes with time. To handle accuracy variance, existing works \cite{jiang_chameleon_2018,li_reducto_2020, zhang_awstream_2018} re-profile using raw video frames periodically to update the accuracy function. Following this method, we profile one frame every 20 seconds and update the accuracy function. We then calculate the distribution of its Mean Absolute Error (MAE) with respect to the ground-truth accuracies of the following 20 seconds in Fig.~\ref{fig-mae-accmap}. As shown in the plot, in \textbf{50\%} cases, the MAE surpasses \textbf{0.05}. As a comparison, the average mIoU difference of our adjacent model version is \textbf{0.042}, indicating that the accuracy estimation error starts to impact the model selection. As such, it suggests that \textit{we need to handle the accuracy fluctuation at segment granularity}.
\subsubsection{Impact on the Selection of Compression Settings}

Before streaming video frames to the edge/cloud server, we first need to decide the compression settings for each segment to satisfy the bandwidth constraint. Unfortunately, the encoded segments' bitrate is not solely decided by the compression settings but also by the video contents. Fig.~\ref{fig-bandwidth-var} shows the variation of segment bitrate of all possible compression settings. It can be seen that the actual bitrate varies drastically across segments, which indicates \textit{we need to consider the impact of varying video contents when we estimate the segment bitrate.}

We further observe a trade-off between QP and resolution. First, we evaluate their efficiency in trading accuracy for bandwidth saving. Specifically, we measured each segment's bitrate efficiencies, i.e., the ratio between bitrate saving and accuracy drop when tuning to a lower quality. Higher bitrate efficiency implies that the knob can trade less accuracy for the same bandwidth saving. Interestingly, as shown in Fig.~\ref{fig-comp-effi}, \textit{QP generally outperforms resolution, proving its efficiency in compressing video volumes with minor accuracy drop}. We also noted significant performance fluctuations on both knobs, indicating a substantial impact from video content. Intuitive thought is that we can always choose to degrade QP instead of a resolution to preserve accuracy as much as possible. However, the \textit{degrading resolution provides an additional opportunity to reduce the inference cost}, as shown in Fig.~\ref{fig-flops-rs}. As such, the trade-off between QP and resolution offers additional room for optimization.

\section{Design}\label{design}
\subsection{System Overview}
Fig.~\ref{fig-se-illu-avc} shows the overall design of \sysname{}. The proposed system has three major components:

\textbf{Bitrate Estimator.} When a new raw video segment is ready, the bitrate estimator predicts its bandwidth usage under all compression settings. Moreover, the bitrate estimator provides information about the current scene dynamics to the CRL adapter. 

\textbf{Performance Encoder.} On the server side, the performance encoder utilizes the softmax probabilities to generate a performance embedding, which is then fed back to the IoT client to aid the decision-making process of the CRL adapter. 

\textbf{CRL Adapter.} The CRL adapter is a DRL model that utilizes estimated bandwidth usage, historical configurations, and performance embedding to choose the compression settings and edge model version for the upcoming video segment. This selection is based on minimizing the inference cost while maintaining target accuracy within the constraints of the available bandwidth.

\subsection{Bitrate Estimator}\label{chp-se}
In this section, we revisit the prediction scheme of H.264/AVC codec. Then we elaborate on the design of the proposed bitrate estimator and training method.

\subsubsection{Dive into H.264/ACV Prediction Scheme}\label{se-h264-dive}

\begin{figure}[t]
    \setlength{\abovecaptionskip}{0.2cm}
	\setlength{\belowcaptionskip}{-0.2cm}
	\centering
	\subfigure[Inter-prediction]{
		\includegraphics[width=0.6\linewidth]{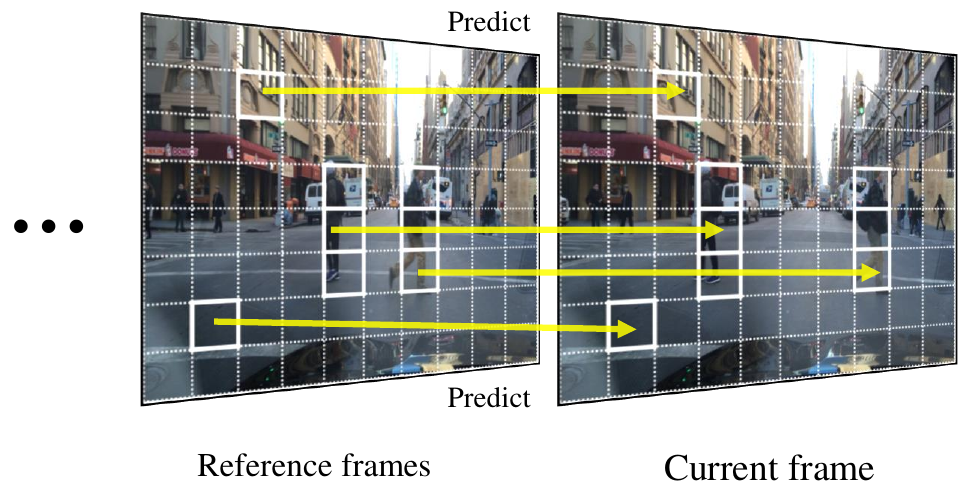}\label{fig3.2.1a}}
	\hfill
	\subfigure[Intra-prediction]{
		\includegraphics[width=0.3\linewidth]{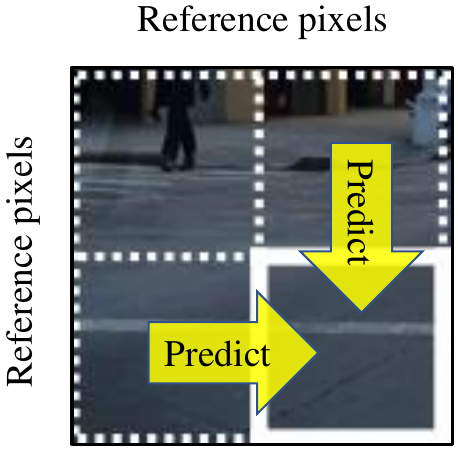}\label{fig3.2.1b}}
        \vspace{-0.2cm}
	\caption{Illustrations of H.264/AVC prediction scheme.}
	\label{fig-se-illus-prediction} 
        \vspace{-0.2cm}
\end{figure}

We start by reviewing the encoding scheme of H.264/ACV. The encoder processes video frames in units of \textit{macroblocks} consisting of 16x16 or 4x4 pixels. It forms the basic units of the \textit{prediction} scheme of H.264/ACV, which exploits the content redundancy to save bitrates. As illustrated in Fig.~\ref{fig-se-illus-prediction}, H.264/AVC adopts two types of prediction: \textit{intra-prediction} and \textit{inter-prediction}. Intra-prediction is used to predict a block of pixels within the same frame. It exploits the spatial correlation between neighboring pixels within the same frame to predict the current block. Inter-prediction is used to predict a block of pixels from a previously encoded frame and uses \textit{motion estimation} to find a block of pixels in a previously encoded frame similar to the current block. Once the reference list between macroblocks is constructed, the encoder subtracts the prediction from the current macroblock to form a residual, which consumes much fewer bitrates after quantization and entropy encoding. 

\begin{figure}[t]
    \setlength{\abovecaptionskip}{0.2cm}
	\setlength{\belowcaptionskip}{-0.2cm}
	\begin{minipage}{0.49\linewidth}
		\centering
		\includegraphics[width=1.0\linewidth]{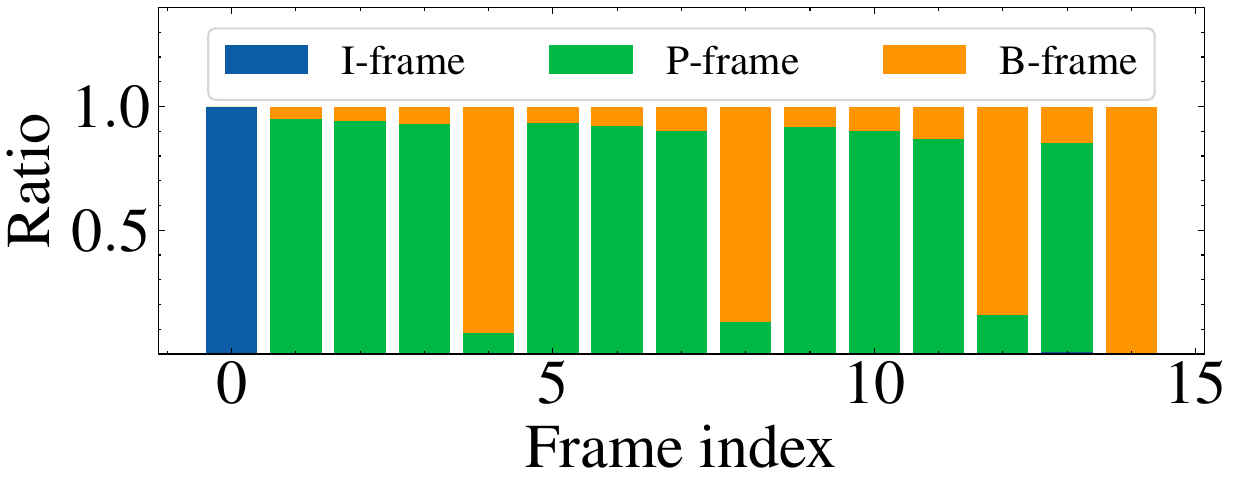}
		\caption{Distribution of frame types.}
		\label{fig-se-frametype}
	\end{minipage}\enspace	
	\begin{minipage}{0.49\linewidth}
		\centering
		\includegraphics[width=1.0\linewidth]{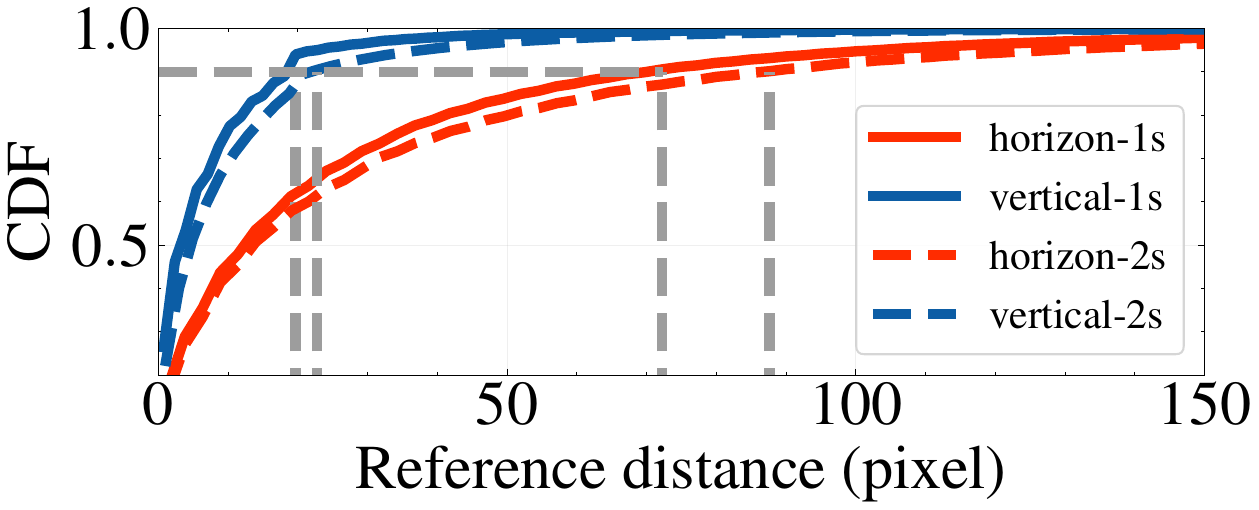}
		\caption{Distribution of motion range.}
		\label{fig-se-merange}
	\end{minipage}\enspace	
        \vspace{-0.4cm}
\end{figure}

Above them, video frames can be categorized into\textit{ I-frames, P-frames, and B-frames}. I-frames contain only macroblocks encoded with intra-prediction, while P-frames and B-frames consist of both macroblocks with intra-prediction and inter-prediction. Accordingly, in Fig.~\ref{fig-se-frametype}, we plot the distribution of frame types for each frame position with video segments encoded in \S\ref{finding-dataset}. It shows that each frame position has a unique frame type distribution. It indicates that \textit{each frame position has a different status in the H.264/AVC prediction scheme.} For instance, the first frame is always an I-frame, which means all of its macroblocks use intra-prediction. As opposed, the tailing frame is always B-frame, which suggests it mainly consists of macroblocks with inter-prediction. 

We further investigate the motion estimation scheme, which involves searching for the best match within a search window around the current macroblock. Typically, the search window size is limited by a max iteration time \cite{noauthor_encodermec_2023} for efficiency. We plot the distribution of the motion range in the x and y axis of both 1-second and 2-second segments in Fig.~\ref{fig-se-merange}. We can observe that the maximum motion range for both segment lengths is below \textbf{100} pixels in \textbf{95\%} of the cases. It indicates that instead of adopting a deep model with a large reception field that covers the full picture (1280x720), \textit{a relatively shallow network with a smaller reception field is sufficient for capturing the bitrate fluctuation brought by video motion}.

\subsubsection{Model Design}

\begin{figure}[t]
    \setlength{\abovecaptionskip}{0.2cm}
	\setlength{\belowcaptionskip}{-0.2cm}
	\centering
	\includegraphics[width=0.9\linewidth]{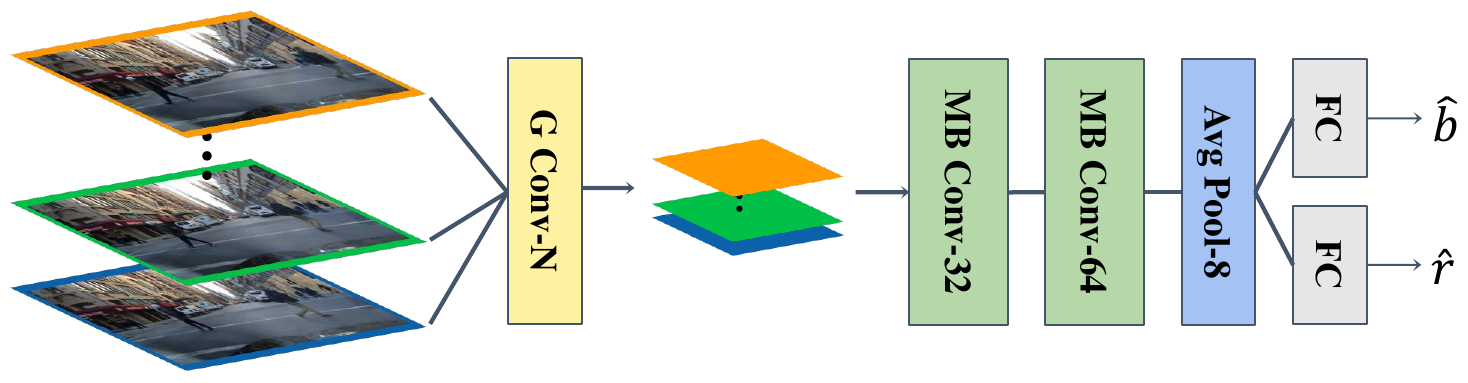}
	\caption{Architecture of the bitrate estimator}
	\Description{}
	\label{fig-se-arch-policy}
        \vspace{-0.4cm}
\end{figure}

Inspired by the above observations, we propose a lightweight model based on the convolutional neural network (CNN) that accurately predicts segment bandwidth usage with raw video frames. As depicted in Fig.~\ref{fig-se-arch-policy}, the bitrate estimator takes in the raw video frames down-sampled by half and predicts the segment bitrate under all compression settings. Our bitrate estimator has two unique designs. First, instead of using expensive 3D convolution or sliding a single 2D convolution layer across all frames, we use a group convolution layer (G Conv-N in Fig.~\ref{fig-se-arch-policy}, where N is the segment frame number) that \textit{assigns dedicated weights to each frame} as the input layer. This design exploits the unique status of each frame position as mentioned in \S\ref{se-h264-dive}. The dedicated convolution kernel weights make learning distinct features from each frame position possible. Second, following the input layer are two MobileNet blocks \cite{sandler2018mobilenetv2}, where we additionally use dilated convolution on the depth-wise convolution layers to expand the reception field. Then, an average pooling layer with a kernel size of 8 summarizes the features. Eventually, the output feature of the feature extractor has a receptive field of 166 pixels on the original video frame, sufficient to cover the motion estimation searching window. 

The predict header is a stack of two fully connected (FC) layers that is further divided into two branches. The first branch predicts the \textit{base bitrate} $\hat{b}$, i.e., the bitrate of the most expensive compression setting. The other branch predicts $\hat{r}$, the ratio between bitrates of the rest configurations settings with respect to the base bitrate.

\subsubsection{Model Training}
As the prediction of segment bandwidth usage is a regression problem, we use standard Mean Square Error (MSE) to train the model:

\begin{small}
	\begin{equation}
		\begin{aligned}
			\theta_{BE} = \arg\operatorname*{min}_{\theta}\frac{1}{N(1+M)}\sum_{i=0}^{N-1}\Big[( b_i - \hat{b}_i )^{2} + \sum_{j=0}^{M-1}( r_{ij} - \hat{r}_{ij} )^{2}\Big]
		\end{aligned}
		\label{eq-se-loss}
	\end{equation}
\end{small}
Where $b_i$ and $r_{ij}$ are the ground-truth base size and size ratio of segment $i$ under compression setting $j$, $\hat{b}_i$ and $\hat{r}_{ij}$ are the corresponding network prediction, respectively.

\subsection{Performance Encoder}\label{chp-pe}
Knowledge about the current segmentation performance is crucial for configuration adaptation in the runtime. Recent work \cite{murad_dao_2022} designed for object detection leverages a YOLO-based \cite{redmon2018yolov3} CNN to predict the runtime accuracy function taking the raw video frames. However, unlike object detection, a sophisticated CNN feature extractor is required to obtain a semantic understanding of the current scenes, which cannot be deployed on general IoT devices only equipped with CPUs. Instead, \textit{we shift this burden to the edge server by extracting features about runtime model performances directly from the edge model outputs of the last available prediction. } 

\begin{figure}[t]
    \setlength{\abovecaptionskip}{0.2cm}
	\setlength{\belowcaptionskip}{-0.2cm}
	\centering
	\includegraphics[width=0.9\linewidth]{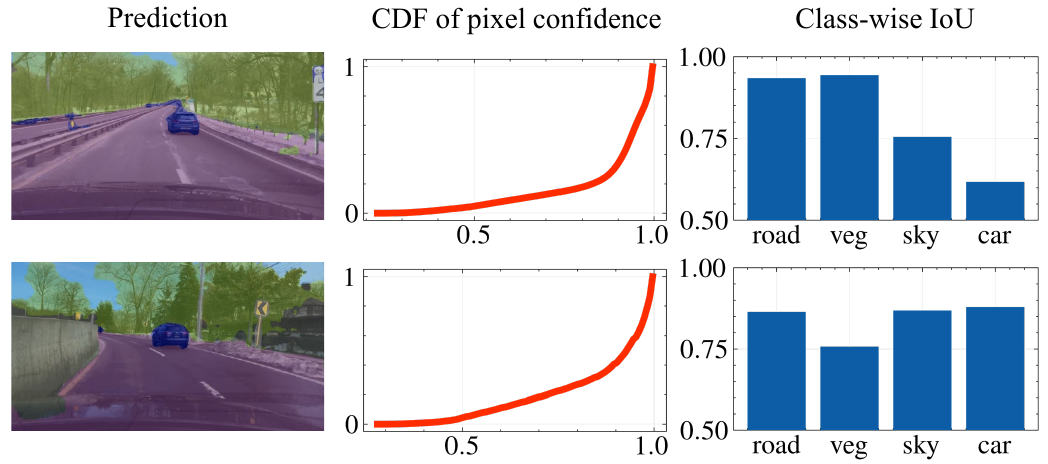}
	\caption{Example demonstrates two frames with comparable pixel confidence distributions but significantly different mean Intersection over Union (mIoU) scores.}
	\Description{}
	\label{fig-pe.1}
\end{figure}

An intuitive approach to implement the performance encoder is directly analyzing the pixel-level prediction confidence, i.e., maximum softmax probability. While this method works for classification \cite{hendrycks_baseline_2018}, it does not correlate well with the mIoU of VSS predictions. As shown in Fig.~\ref{fig-pe.1}, the upper frame has a mIoU of 0.63 while the lower frame has a mIoU of 0.80, even though they have similar pixel confidence distributions. This is because unlike pixel accuracy, which treats every pixel equally, \textit{mIoU intrinsically weights each pixel according to the number of pixels within the class it belongs to}. 
The right side of Fig.~\ref{fig-pe.1} visually represents class IoU for several classes. It can be observed that the IoU of the "car" and "sky" classes significantly influence the overall mean IoU even though they contain fewer pixels than the class "road".

\begin{figure}[t]
    \setlength{\abovecaptionskip}{0.2cm}
	\setlength{\belowcaptionskip}{-0.2cm}
	\centering
	\includegraphics[width=0.9\linewidth]{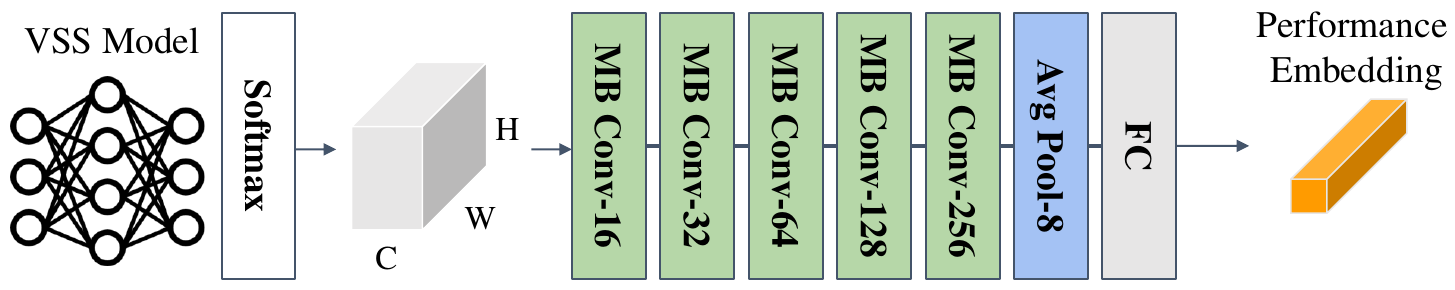}
	\caption{Architecture of the performance encoder}
	\Description{}
	\label{fig-pe-arch}
        \vspace{-0.4cm}
\end{figure}

As such, instead of solely analyzing the pixel confidence, the performance encoder takes \textit{softmax probabilities of all classes} as its input, as shown in Fig.~\ref{fig-pe-arch}. The rationale behind this is threefold.
(\romannumeral1) It provides class information that affects the final mIoU, as mentioned above.
(\romannumeral2) It implies the object sizes in the scenes which helps to explore the opportunity of degrading resolution.
(\romannumeral3) It offers the encoder rich information about the candidate classes whose probability is only second to the largest. For example, if the difference between the probabilities of the first and second classes is small, it may be useful to consider both classes as possible predictions.
Note that our method regards the edge VSS model as a black box; it only needs the final softmax probability of model outputs. Thus it can be applied to various VSS architectures without further modifications.  

We build the performance encoder with 5 MobileNet blocks followed by an average pooling layer as illustrated in Fig.~\ref{fig-pe-arch}. We rescaled the probabilities to Cx320x180 for all resolutions (C is the number of classes), leading to a low inference cost $<$ 0.5G FLOPs. The flattened extracted features will be transmitted back to the IoT device as \textit{performance embedding} to aid the configuration adaption process. This design leverages the rich GPU resource on the edge server and avoids imposing demanding feature extraction on IoT devices. The encoder is optimized end-to-end along with the CRL adapter, which will be introduced in section \S\ref{chp-adapter}.

\subsection{CRL Adapter}\label{chp-adapter}
In this section, we elaborate on the design of the CRL Adapter. We first formulate the problem theoretically and then describe the details of our solution.

\subsubsection{Problem Formulation}\label{chp-adapter-problem}
In \sysname{}'s framework, the captured raw video frames are encoded and streamed in segments of $T$ seconds, which is the basic unit of configuration adaptation. When segment $i$ is ready, the goal of the CRL Adapter is to choose the best configuration that minimizes the edge inference cost of segment $i$ under the constraints of bandwidth and accuracy. We formulate it into the following problem:

\begin{small}
\begin{equation}
\setlength{\abovedisplayskip}{0pt}
\setlength{\belowdisplayskip}{0pt}
	\begin{aligned}
			& \operatorname*{min}_{r,q,v} C(r,v)\\
	  \text{s.t.,} \  \   & b_i(r,q)\le B_i\\
			& \lambda_i(r,q,v) \ge A_i\\
		\end{aligned}
\label{eq3.1}
\end{equation}
\end{small}

Where $C$ is the cost function. $b_i$ and $\lambda_i$ are the segment bitrate and average mIoU across frames in segment $i$, respectively. $B_i$ is the bandwidth of the following $T$ second at the uploading time of segment $i$, and $A_i$ is the target accuracy set by the user. $r$, $q$, $v$ is the selected resolution, QP, and edge model version, respectively. If the device fails to upload segment $i$ in $T$ seconds, it will drop segment $i$ and start to stream segment $i+1$ instead. This strategy avoids draining the on-device video buffer and limits the queuing delay \cite{murad_dao_2022}. The accuracy of the incomplete segment is zero, as the server cannot decode it. 

In some extreme cases, there is no solution to problem \eqref{eq3.1}. For instance, the target accuracy cannot be achieved when the bandwidth is too low. In this situation, we hope to optimize the following problem instead:

\begin{small}
\begin{equation}
\setlength{\abovedisplayskip}{0pt}
\setlength{\belowdisplayskip}{0pt}
	\begin{aligned}
			& \operatorname*{max}_{r,q,v} \lambda_i(r,q,v)\\
	  \text{s.t.,} \  \   & b_i(r,q)\le B_i\\
		\end{aligned}
	\label{eq3.2}
\end{equation}
\end{small}

It will maximize the inference accuracy under the constraint of bandwidth instead of minimizing inference cost to ensure the quality of serving. 

\subsubsection{State Space Design}
In DRL, the state space represents the environmental information that the policy can perceive at step $i$. 
In our solution, the state at step $i$ is defined as:

\begin{small}
\begin{equation}
	s_i = (A_i, B_i , b_i, p_{i-1}, a_{i-1} , x_i)
	\label{eq3.3}
\end{equation}
\end{small}

Where $A_i$ and $B_i$ are the accuracy and bandwidth constraints. $b_i$ is the estimated bitrate of segment $i$. $p_{i-1}$ is the performance embedding extracted from the last frame of the latest predicted segment (segment $i-1$, for instance).  
$a_{i-1}$ is the corresponding configuration of segment $i-1$. It represents the resolution, QP, and edge model version set to segment $i-1$. Including this term makes it resistant to the dynamic of final performance encoding caused by different configurations.  Moreover, as the performance embedding is extracted from the last frame of segment $i-1$, its effectiveness in the decision of segment $i$ depends on how much the video contents have changed in segment $i$. Thus, we further append $x_i$, the features before decision layers of the bitrate estimator, into the state space. As described in \S\ref{chp-se}, the bitrate estimator captures the content dynamics of segment $i$, which makes the feature suitable as hints to inform the policy of the current content dynamics.

\subsubsection{Policy Design}
The policy is the core component of reinforcement learning. At step $i$, the policy $\pi$ receives the current state $s_i$ and produces corresponding action $a_i$. 
A policy can be either deterministic or stochastic. A deterministic policy map states a single action, while a stochastic policy map states a probability distribution over actions. 
We build a stochastic policy for its advantage in highly dynamic environments with imperfect information \cite{sutton2018reinforcement}. 
Specifically, the policy maps $s_i$ to $a_i$, a probability distribution of all possible configurations. 
The action is sampled from the distribution during the training phase, and the action with the highest probability is selected during the testing phase. 

\begin{figure}[t]
    \setlength{\abovecaptionskip}{0.2cm}
	\setlength{\belowcaptionskip}{-0.2cm}
	\centering
	\includegraphics[width=\linewidth]{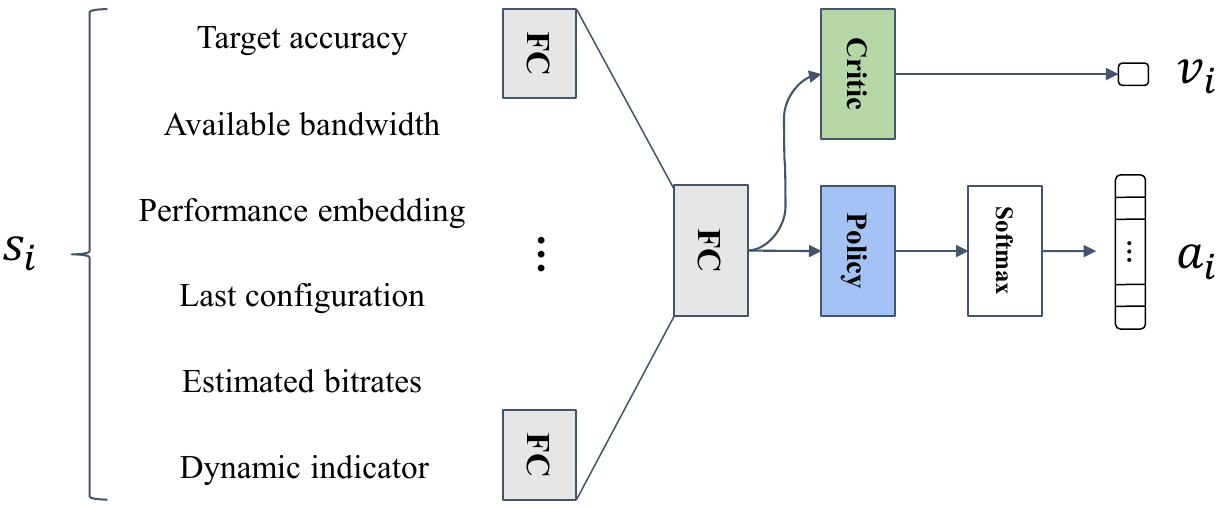}
	\caption{Architecture of the CRL adapter}
	\Description{}
	\label{fig-rl-arch-policy}
        \vspace{-0.4cm}
\end{figure}

Due to the high-dimensional continuous state space, we leverage the powerful approximation ability of neural networks and represent our policy with it. As shown in Fig.~\ref{fig-rl-arch-policy} we build the policy network with FC layers. Specifically, we first use separate FC layers to extract local features, which are further concatenated and fed to the following layers. The output of the policy network is a vector representing the probability of all configurations.

\subsubsection{Reward Design with Constraints}
At step, $i$, the predicted $a_i$ configures the segment $i$. A corresponding reward $r_i$ is generated by the reward function $R$, which evaluates the advantage of the move $a_i$ on $s_i$. The rewards will be used as supervisory signals to optimize the policy network. In standard reinforcement learning settings, the optimization goal of the policy network is to maximize the expected sum of discounted rewards $J(\pi)$:

\begin{small}
\begin{equation}
\setlength{\abovedisplayskip}{0pt}
\setlength{\belowdisplayskip}{0pt}
	\operatorname*{max}_{\pi} J(\pi)=\operatorname*{max}_{a\sim\pi,s\sim P}\left[\sum_{i}\gamma^{i}r_{i}\right]
	\label{eq3.4}
\end{equation}
\end{small}
where $P$ is probability of transitioning to state $s^{i+1}$ from $s_i$ when an action $a_i$ is taken. $\gamma$ is the reward discount factor and $r_i$ is the reward on taking $a_i$ given $s_i$. However, considering the constraints in \eqref{eq3.1} and \eqref{eq3.2}, we need to extent Eq.~\eqref{eq3.4} to the new form of constrained reinforcement learning:
\begin{small}
\begin{equation}
\setlength{\abovedisplayskip}{0pt}
\setlength{\belowdisplayskip}{0pt}
	\begin{aligned}
		&\operatorname*{max}_{\pi} J(\pi) \\ 
		\text{s.t.,} \  \   &J^{c_i}(\pi)\le\beta_i\ \forall i\\
	\end{aligned}
	\label{eq3.5}
\end{equation}
\end{small}

where $J^{c_i}$ and $\beta_i$ represent the constraints to be met at each step $i$. A common way to solve Eq.~\eqref{eq3.4} is to solve its Lagrangian relaxation instead \cite{calian2020balancing, paternain2019constrained}. However, these methods require careful tuning of the Lagrangian parameters and are hard to optimize for complex tasks \cite{mandhane_muzero_2022}. We thus extend the self-competition reward scheme in \cite{mandhane_muzero_2022} to solve problem \eqref{eq3.1} and \eqref{eq3.2} and define the reward at step $i$ as the following:

\newcommand{\EMAcost}{{C_{EMA}}}
\newcommand{\EMAbit}{{O_{EMA}}}
\newcommand{\EMAacc}{{U_{EMA}}}
\newcommand{\nowcost}{{C_{i}}}
\newcommand{\nowbit}{{O_{i}}}
\newcommand{\nowacc}{{U_{i}}}
\newcommand{\constrainindicator}{{\delta_{i}}}
\newcommand{\constrainindicatorinv}{{\xi_{i}}}

\begin{small}
\begin{equation}
\setlength{\abovedisplayskip}{0pt}
\setlength{\belowdisplayskip}{0pt}
\begin{aligned}
	&r_i = \mathrm{{sgn}}(\EMAcost-\nowcost)\constrainindicator - \constrainindicatorinv \\
\end{aligned}
\label{eq3.6}
\end{equation}
\end{small}
where
\begin{small}
\begin{equation}
\begin{aligned}
&\constrainindicator=\mathbbm{1}_{\EMAacc\le0,\EMAbit\le0,\nowbit\le0,\nowacc\le0} \\
&\constrainindicatorinv=\max \{\mathbbm{1}_{\nowacc\ge\EMAacc},\mathbbm{1}_{\nowbit\ge\EMAbit}\}
\end{aligned}
\label{eq3.6.1}
\end{equation}
\end{small}

$\nowcost$ is the server inference cost. $\nowacc$ and $\nowbit$ are $A_i-\lambda_i$ and $b_i-B_i$ that indicates if the constraints are met. It also maintains their exponential moving averages (EMA) 
as $\EMAacc$, $\EMAbit$, and $\EMAcost$, respectively. 

The intuition behind Eq.~\eqref{eq3.6} is to force the policy beat its previous behaviors based on EMA. Specifically, $\constrainindicator$ judges if the accuracy and bandwidth constraints are satisfied. If so, the first term in Eq.~\eqref{eq3.6} will minimize the inference cost by beating its historical performance $\EMAcost$. Otherwise, the first term will be ignored, and the second term $\constrainindicatorinv$ will punish the policy if any constraints are violated. As such, Eq.~\eqref{eq3.6} only allows the policy to optimize for inference cost when the constraints are consistently satisfied.

\subsubsection{Policy Optimization}
\newcommand{\valuefunc}{{V_{\phi}}}
We have defined the state $s_t$, action $a_t$, and reward $r_t$. The next step is to optimize the policy. First, the policy will act as the CRL Adapter in \sysname{} and roll out trajectories, i.e., states, actions, and rewards sequences. Then we use the actor-critic approach \cite{mnih2016asynchronous} to calculate the advantage of actions. As shown in Fig.~\ref{fig-rl-arch-policy}, a parameterized value function $\valuefunc$ named critic network, which shares the same architecture with the policy network, is used as a baseline. It takes the same input as the policy network and predicts $v_i=\valuefunc(s_i)$ as the expected reward of $s_i$. The advantage of step $i$ is then calculated as $A_i = r_i - v_i$, and it evaluates how good the action is compared to the average expected reward under $s_i$. This design reduces variance in the policy gradient and stabilizes the training process. $A_i$ essentially replace $r_i$ in Eq.~\eqref{eq3.4}.

Given the advantages, we then optimize the policy network and $V_{\theta_v}$ with Proximal Policy Optimization (PPO) algorithm \cite{schulman_proximal_2017} for its sample efficiency and stability. Different from vanilla policy gradient, PPO updates the policy network multiple times with the same batch of collected trajectories using the clipped surrogate objective function:

\newcommand{\probratio}{{R_i(\theta_{k},\theta)}}

\begin{small}
\begin{equation}
\setlength{\abovedisplayskip}{0pt}
\setlength{\belowdisplayskip}{0pt}
	\begin{aligned}
		L^{policy}(\theta_{k},\theta)&=\operatorname*{min}\Big(\probratio A_i, \operatorname{clip}\Big(\probratio ,1-\epsilon,1+\epsilon\Big)A_i\Big) \\
	\end{aligned}
	\label{eq3.7}
\end{equation}
\end{small}
where
\begin{small}
\begin{equation}
\setlength{\abovedisplayskip}{0pt}
\setlength{\belowdisplayskip}{0pt}
	\begin{aligned}
		\probratio &= \frac{\pi_{\theta}(a_i|s_i)}{\pi_{\theta_{k}}(a_i|s_i)}
	\end{aligned}
	\label{eq3.7.1}
\end{equation}
\end{small}

$\theta_k$ is the old policy parameter before updated, $\probratio$ is the probability ratio of action $a_i$ between old and new policy given the same state $s_i$. Through the clipping operation, Eq.~\ref{eq3.7} carefully updates the policy $\pi_{\theta}$ while controlling the step by avoiding the behavior of the new policy getting too far from where it starts. Here $\epsilon$ is a hyper-parameter that controls the bound of clipping. Then the policy parameters are updated by maximizing the objective function on a collection of trajectories:

\begin{small}
\begin{equation}
\setlength{\abovedisplayskip}{0pt}
\setlength{\belowdisplayskip}{0pt}
	\begin{aligned}
	\theta = \arg\operatorname*{max}_{\theta}\sum_{i}L^{policy}
	\end{aligned}
	\label{eq3.8}
\end{equation}
\end{small}

At the same time, the value function $\valuefunc$ is fitted by regression on the reward $r_i$:

\begin{small}
\begin{equation}
\setlength{\abovedisplayskip}{0pt}
\setlength{\belowdisplayskip}{0pt}
	\begin{aligned}
		\phi=\mathrm{arg}\operatorname*{min}_{\phi}\sum_{i}\,\Big(\valuefunc(s_i)-r_i\Big)^{2}
	\end{aligned}
	\label{eq3.9}
\end{equation}
\end{small}

\section{Evaluation}\label{eval}

\subsection{Experiment Setup}
\subsubsection{Dataset and Model Training}\label{exp-dataset&train}
We randomly select 300 videos totaling 200 minutes to build our training dataset. Following the procedure in \S\ref{finding-dataset}, we split the videos into 1-second segments with a frame rate of 15 fps and compressed them using H.264/AVC using 18 different settings. A test set of 60 random videos totaling 40 minutes is built and processed with the same procedure.

We first trained the bitrate estimator. We used a learning rate that gradually reduced from 1e-4 to 1e-6 through cosine annealing, set the weight decay to 1e-4, and continued with a 200-epoch training process. Subsequently, we proceeded to jointly train the frozen bitrate estimator, the CRL adapter, and the performance encoder. At the beginning of each epoch, we uniformly sampled the target accuracy and available bandwidth from $\{0.5,0.55,\ldots,0.75,0.80\}$ and $[0.3, 1.0]$ MBps, respectively. We set the batch size to 32, $\epsilon$ to 0.2, and $\gamma$ to 0.9, which means every action will be evaluated by the performance of the future ten segments. We set the exponential smoothing constant $\alpha$ to calculate EMA to 0.2. The learning rates of the policy and value function are both 1e-4. We trained the policy for 2 million steps.

\subsubsection{Implementation}
The edge server is an Ubuntu workstation equipped with Intel Xeon Gold 6226R CPU, 128 GB RAM, and a Geforce RTX 4090 GPU. We use a Raspberry Pi 4B \cite{ltd_raspberry_nodate} as the IoT device that embeds a 1.8 GHz quad-core CPU and 4 GB of RAM. We recompile the H.264 codec in FFmpeg \cite{tomar2006converting} to work with the onboard video-core to accelerate video encoding.  Note that the onboard video-core is specially prepared for the codec, and all computations of \sysname{} are conducted using CPUs only. We use the Traffic Control (TC) tool of Linux to control the bandwidth between IoT devices and the edge server. 

\subsubsection{Baselines}
We compare \sysname{} with the following baselines:

\newcommand{\casvaname}{{\textit{CASVA-C}}}
\newcommand{\daoname}{{\textit{DAO-C}}}
\newcommand{\reprofilename}{{\textit{Periodic Profiling}}}
\newcommand{\optname}{{\textit{Optimal}}}

\textbf{Periodically Profiling}:
This method streams the first frame of each 40-second video to the server and profiles the accuracies of all 90 configurations. Then it selects the best configuration for the rest of the segments through an exhaustive search. We let it access the ground-truth segment bitrate to cope with the bandwidth constraints.

\textbf{CASVA-C}:
CASVA \cite{zhang_casva_2022} is a DRL video analytic framework that maximizes inference accuracy under bandwidth constraints. It encodes each segment one more time and uses its bitrate as an indicator of video content dynamics. We modified it to cope with our cost-reduction task. It is similar to \sysname{} but replaces the performance encoder with the method above. We named this approach as \casvaname{} and trained it following the same procedure in \S\ref{exp-dataset&train}.

\textbf{DAO-C}:
DAO \cite{murad_dao_2022} uses a YOLO \cite{redmon2018yolov3} based accuracy estimator to predict runtime accuracy function with the first frame of each segment. To cope with our settings, we instead trained a MobileNet-based semantic segmentation network \cite{zhang_fast_2023} on the training set of BDD100k. Then we train the accuracy estimator with transfer learning following the procedure in \cite{murad_dao_2022}. At runtime, it uses the same strategy of \reprofilename{} to find the best configuration.

\textbf{Optimal}:
A perfect model can access every segment's ground-truth accuracy and bitrate. It always chooses the optimal configuration and represents the upper bound of \sysname{}'s performance.

\subsection{Evaluation Results}
\begin{figure}[t]
    \setlength{\abovecaptionskip}{0cm}
	\setlength{\belowcaptionskip}{-0.4cm}
	\centering
	\subfigure{
		\includegraphics[width=\linewidth]{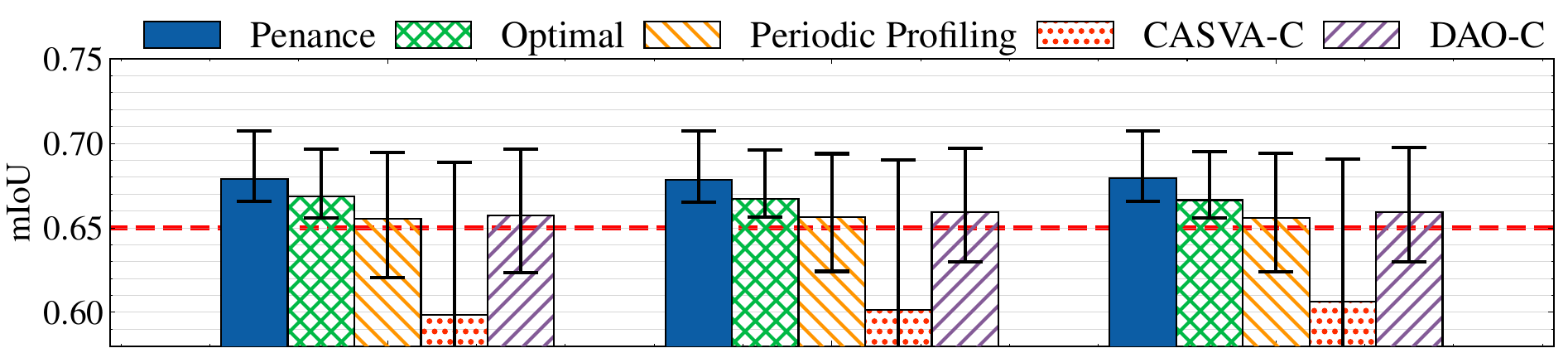}\label{fig-eval-fixaccvarbw-acc}}
	\hfill
        \vspace{-0.5cm}
	\subfigure{
		\includegraphics[width=\linewidth]{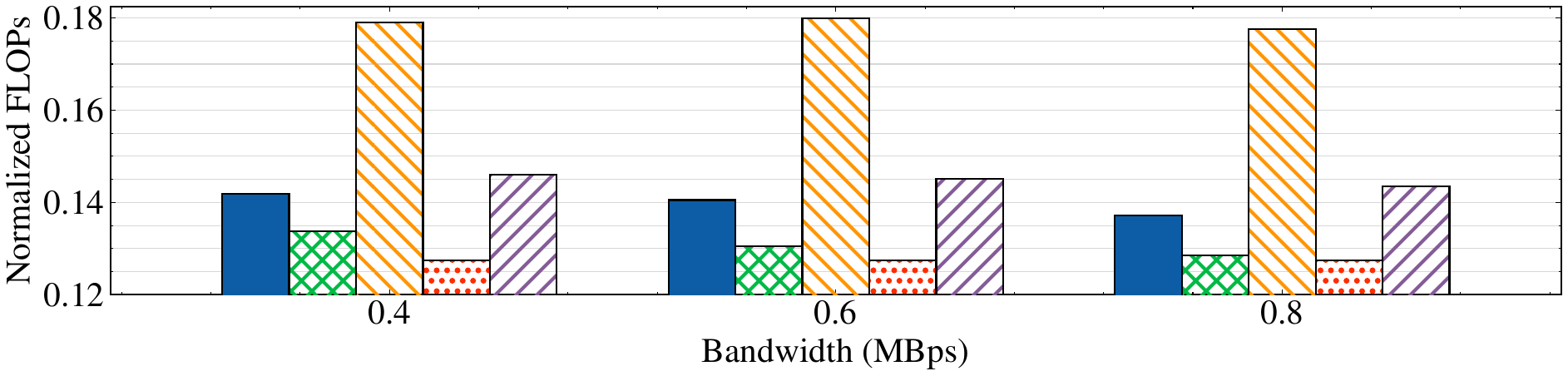}\label{fig-eval-fixaccvarbw-cost}}
        \vspace{-0.2cm}
	\caption{Comparing \sysname{} with baseline methods under different bandwidths with target mIoU of 0.65.}
	\label{fig-eval-fixaccvarbw} 
\end{figure}

\subsubsection{Overall Performance}
We first investigate the overall performance of \sysname{} by fixing the target mIoU to 0.65 and analyzing the achieved mIoU under bandwidth ranging from 0.4 MBps to 0.8 Mbps. We plot the results in Fig.~\ref{fig-eval-fixaccvarbw-acc} where the bars report the \textit{25th} percentiles, and the error bars show the \textit{median} and \textit{10th} to better describe the skewed accuracy distribution. As demonstrated, \sysname{} is able to achieve the target accuracy under different bandwidth conditions, with an average failure rate (averaged ratio of segments that fail to achieve the target accuracy across bandwidths) of \textbf{4.1\%}. Although the median accuracy of \reprofilename{} and \daoname{} surpass the target accuracy, they failed in 36.2\% and 34.4\% cases, respectively, due to the inaccurate accuracy map estimation. While \casvaname{} use the same DRL framework as \sysname{}, it has a large average failure rate of 60.5\%. This result suggests that the correlation between encoded segment size and runtime mIoU is relatively weak. While the segmentation difficulty of the video frames only relates to the current frame contents, the encoded segment size is additionally affected by the inter-frame similarity as illustrated in \ref{se-h264-dive}, making it noisy to represent the frame complexity.

Despite the accuracy assurance, \sysname{} additionally consumes only \textbf{6.8\%} FLOPs compared with \optname{} as illustrated in Fig.~\ref{fig-eval-fixaccvarbw-cost}, where the bars represent the averaged segment FLOPs normalized by the cost of the most expensive model with full resolution inputs. Though only one frame is profiled for every 40-second video, \reprofilename{} is still very expensive and spends 36.7\% more FLOPs than the \optname{}. While \casvaname{} instead consumes fewer FLOPs than \optname{}, it violates the accuracy constraint. Though \daoname{} performs slightly better than the other two baselines in maintaining accuracy, it consumes 10.7\% more FLOPs than \optname{}. It validates the effectiveness of the feedback design of our performance encoder. We can also observe that the FLOPs of \sysname{} degrade with the increasing bandwidth, indicating the internal balancing between bandwidth consumption and inference cost reduction.

\begin{figure}[t]
    \setlength{\abovecaptionskip}{0cm}
	\setlength{\belowcaptionskip}{-0.4cm}
	\centering
	\subfigure{
		\includegraphics[width=\linewidth]{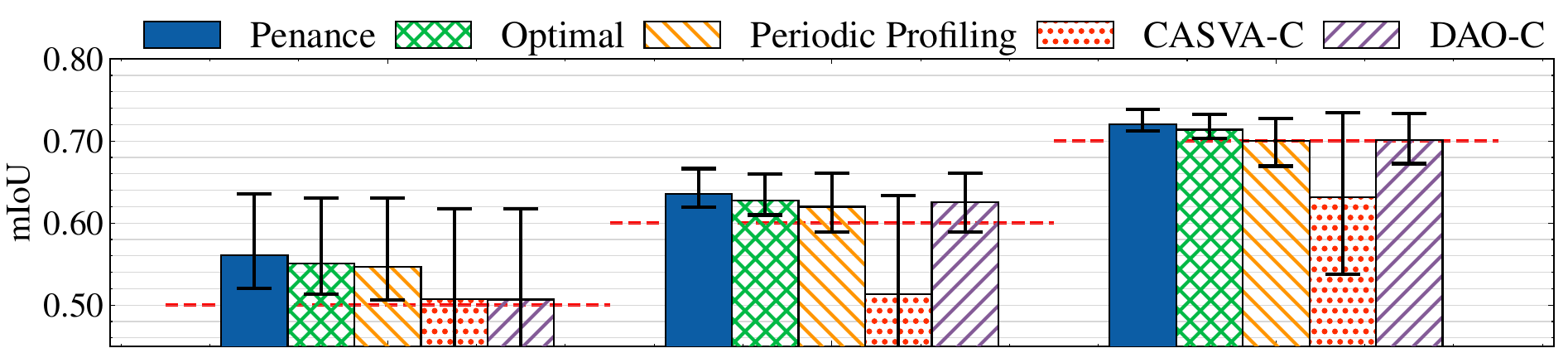}\label{fig-eval-fixbwvaracc-acc}}
	\hfill
        \vspace{-0.5cm}
	\subfigure{
		\includegraphics[width=\linewidth]{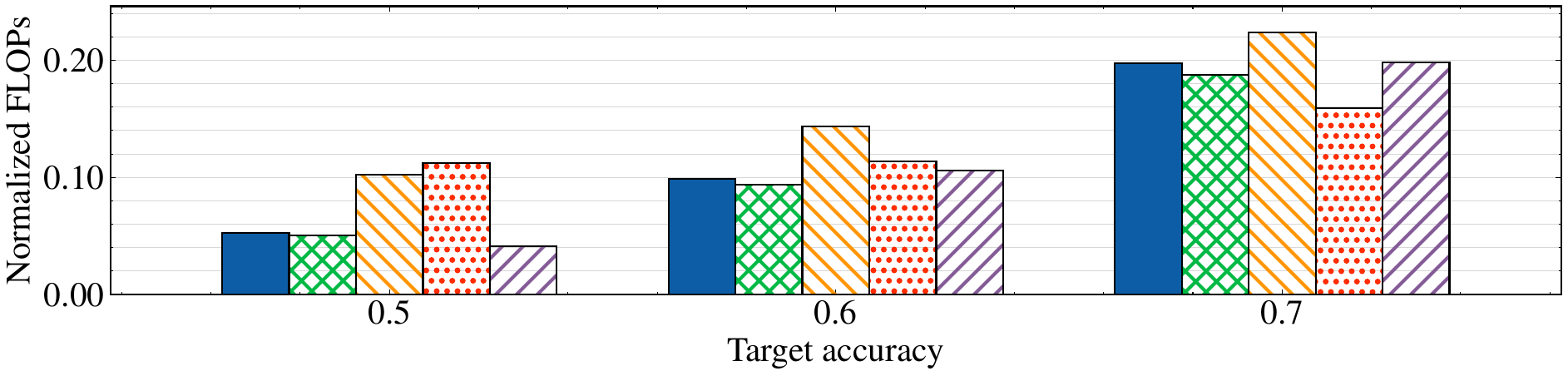}\label{fig-eval-fixbwvaracc-cost}}
        \vspace{-0.2cm}
	\caption{Comparing \sysname{} with baseline methods by varying target mIoU with bandwidth limited to 0.6 MBps.}
	\label{fig-eval-fixbwvaracc} 
\end{figure}
\subsubsection{Varying Target Accuracy}
We further evaluate the performance of \sysname{} given different accuracy requirements. In this experiment, we fix the bandwidth to 0.6 MBps, vary the target mIoU from 0.5 to 0.7, and plot the results in Fig.~\ref{fig-eval-fixbwvaracc-acc}, where the bars report the \textit{25th} percentiles, and the error bars show the \textit{median} and \textit{10th}. As shown in the plot, \sysname{} successfully catches up with the varying target accuracy, with the failure rate of 3.6\%, 3.9\%, and 7.3\%, respectively, while the other baselines either fail to meet the target accuracy or consume a lot more FLOPs than \sysname{}. Interestingly, for the target mIoU of 0.5, \daoname{} reduces the inference cost aggressively, leading to a failure rate of 40.8\%, while \casvaname{} becomes too conservative that it spends 122.9\% more FLOPs compared with \optname{}. This result can be explained by the fact that with the degradation of frame quality and model complexity, the relationship between configurations and runtime edge model accuracy is even harder to predict as the image noise and model imperfection start to exert more influence on the model outputs. On the contrary, by directly analyzing the final output probabilities of edge models, \sysname{} can still effectively monitor the runtime performance.

\begin{table}[t]
    \centering
    \renewcommand{\arraystretch}{0.9}
    \caption{Performance of segment bitrate estimation}
    \vspace{-0.4cm}
    \begin{tabular}{l|l|llcc}
        \toprule
        Setting & Algorithm & Mean & $P_{25\%}$ & Median & $P_{75\%}$ \\
        \midrule
        \multirow{2}{*}{1s-15fps}   & AR & 25.07\% & 16.39\% & 23.71\% & 30.41\%\\
                                    & Ours & \textbf{17.22\%} & \textbf{6.24\%} & \textbf{12.96\%} & \textbf{24.31\%}\\
        \midrule
        \multirow{2}{*}{1s-30fps}   & AR & 28.67\% & 19.39\% & 27.16\% & 35.87\% \\
                                    & Ours & \textbf{16.72\%} & \textbf{6.76\%} & \textbf{12.19\%} & \textbf{22.18\%}\\
        \midrule
        \multirow{2}{*}{2s-15fps}   & AR & 31.25\% & 14.38\% & 24.73\% & 37.19\% \\
                                    & Ours & \textbf{20.54\%} & \textbf{6.50\%} & \textbf{14.00\%} & \textbf{21.77\%}\\
        \bottomrule
    \end{tabular}
    \label{tab-eval-bitrate-estimation}
    \vspace{-0.1cm}
\end{table}

\subsubsection{Bitrate Estimation Performance}
We then evaluate the performance of the bitrate estimator. Here, we compare it with the solution proposed in recent work \cite{liu_adamask_2022}. Specifically, it predicts the base bitrate with an autoregression ($AR$) model and uses an offline-learned ratio table to derive the bitrates of the rest compression settings. We adopt the first-order autoregressive model ($AR(1)$) as it outperforms other order settings and regresses the ratio table on the same training set used by the bitrate estimator. We demonstrate the relative bitrate error statistics in the first row of Table \ref{tab-eval-bitrate-estimation}. The results show that our method outperforms the baseline method. This is because the AR method cannot faithfully forecast the future bitrate when large dynamics exist in the video scenes. Instead, our method predicts per-segment bitrate consumption directly based on the raw video frames aware of the prediction mechanism of H.264/ACV codecs, making it robust to content dynamics.

To further investigate the generalization ability of the bitrate estimator, we additionally train and evaluate its performance by varying fps and segment duration and show the results in Table \ref{tab-eval-bitrate-estimation}. As expected, our method maintains its performance in different encoding settings. Note that even when the segment duration is doubled, which means more scene changes are involved, the performance of our method does not change much. This result is aligned with our observation in \S\ref{chp-se} that our network design is sufficient to handle the bounded motion distance.

\begin{table}[t]
    \setlength{\abovecaptionskip}{0.2cm}
    \centering
    \renewcommand{\arraystretch}{0.9}
    \caption{Overhead of \sysname{} on \iotdevice}
    \vspace{-0.2cm}
    \begin{tabular}{l|ccc}
        \toprule
                     & Bitrate Estimator & CRL Adapter & Total \\
        \midrule
        Runtime (ms)    & 202.44 & 20.69 & 223.13\\
        FLOPs (M)       & 45.79 & 5.50 & 51.29\\  
        \bottomrule
    \end{tabular}
    \label{tab-eval-computation-overhead}
    \vspace{-0.1cm}
\end{table}

\subsubsection{Computation Overhead on the IoT Device}

This section evaluates our IoT device's computation overhead \sysname{}. We measure the runtime and FLOPs of the two core parts in \sysname{}, i.e., bitrate estimator and CRL adapter, on 1s-15 fps segments (the input data shape is 15x640x360). As illustrated in Table \ref{tab-eval-computation-overhead}, \sysname{} takes $\sim$ 50 M FLOPs, and the total runtime is below 300 ms.  Considering that \sysname{} runs once for each 1s segment, it is able to be deployed on general IoT devices.

\section{Related Works}\label{related}

\textbf{Video Semantic Segmentation:} 
The most straightforward method for VSS is to apply image semantic segmentation models to each frame of the videos, as image semantic segmentation \cite{zhao_icnet_2018,zhao_pyramid_2017, yu_bisenet_2018, wang_swiftnet_2021} does. To fully exploit the relationship between frames, another stream of works uses optical flow or other motion cues to propagate information between frames to improve accuracy \cite{vss-acc-chandra2018deep, vss-acc-gadde2017semantic, vss-acc-jin2017video, vss-acc-nilsson2018semantic} or inference speed \cite{vss-cost-hu2020temporally, vss-cost-li2018low, vss-cost-mahasseni2017budget, vss-cost-zhu2017deep}. \sysname{} is orthogonal to these works as it is agnostic to the details of the VSS model. One can easily conjunct \sysname{} with any VSS model and use it as a plug-in to further reduce the inference cost. 

\textbf{Cost Reduction by Prediction Reusing:} 
Much effort has gone into reducing the computation cost of video analytics on edge servers. One stream of works aims to filter frames without harming the overall accuracy of vision tasks. For instance, FilterForward \cite{canel_scaling_2019} and Noscope \cite{kang_noscope_2017} adopt cheap neural detectors to determine whether the frame should be offloaded. Reducto \cite{li_reducto_2020} leverages low-level image features to enable on-device filtering. Infi \cite{yuan_infi_2022} proposes an end-to-end learnable filtering framework. Glimpse \cite{chen_glimpse_2015} propagates the cached bounding box on-device to future frames with optical flow. 
FoggyCache \cite{guo_foggycache_2018} exploits cross-device data similarity between IoT devices to minimize redundant computation.
However, none of these works investigates pixel-level labeling tasks like VSS for severe accuracy degradation \cite{zhang_edge_2022}. While the recent work EdgeIS \cite{zhang_edge_2022} reused the instance segmentation results with contour calibration, it is optimized for Mask-RCNN and can only trace stationary objects. Instead, Penance is a general edge-assisted VSS framework agnostic to the underlying VSS model.

\textbf{Joint Data and Model Adaptation:} In the realm of image classification and object detection, a few works have been proposed to balance between data versions to minimize the inference costs while satisfying accuracy and bandwidth constraints. For instance, JCAB \cite{wang_joint_2020, zhang_adaptive_2022} jointly optimizes bandwidth allocation and object detection model selection. It models the analytic accuracy as a periodically updated function of frame resolution and sampling rate. $A^2$ \cite{jiang_joint_2021} minimizes the inference cost while meeting accuracy and latency requirements with an offline-generated accuracy map. 
Unfortunately, these works adopt fixed or heuristic accuracy functions specially designed for object detection or image classification, which can not be applied to handle the accuracy fluctuation of VSS models.

\section{Conclusion}\label{conclusion}
This paper presents \sysname{}, a lightweight DRL-based low-cost edge-assisted VSS framework that adapts segment-level configurations to minimize edge inference cost while satisfying the accuracy and bandwidth constraints by leveraging output softmax probabilities and H.264/ACV encoding schemes. Experiment results showed the superiority of \sysname{} over baseline methods.

\section{Acknowledgments}

This work was supported in part by the Key R \& D Program of Hubei Province of China under Grant No. 2021EHB002, the National Science Foundation of China with Grant 62071194, Knowledge Innovation Program of Wuhan-Shuguang, the European Union's Horizon 2020 research and innovation programme under the Marie Skłodowska-Curie grant agreement No 101022280, Horizon Europe MSCA programme under grant agreement number 101086228,
EPSRC with RC Grant reference EP/Y027787/1, and Project funded by China Postdoctoral Science Foundation 2023M731196.

\bibliographystyle{ACM-Reference-Format}
\balance
\bibliography{red-base}


\begin{thebibliography}{56}


\ifx \showCODEN    \undefined \def \showCODEN     #1{\unskip}     \fi
\ifx \showDOI      \undefined \def \showDOI       #1{#1}\fi
\ifx \showISBNx    \undefined \def \showISBNx     #1{\unskip}     \fi
\ifx \showISBNxiii \undefined \def \showISBNxiii  #1{\unskip}     \fi
\ifx \showISSN     \undefined \def \showISSN      #1{\unskip}     \fi
\ifx \showLCCN     \undefined \def \showLCCN      #1{\unskip}     \fi
\ifx \shownote     \undefined \def \shownote      #1{#1}          \fi
\ifx \showarticletitle \undefined \def \showarticletitle #1{#1}   \fi
\ifx \showURL      \undefined \def \showURL       {\relax}        \fi
\providecommand\bibfield[2]{#2}
\providecommand\bibinfo[2]{#2}
\providecommand\natexlab[1]{#1}
\providecommand\showeprint[2][]{arXiv:#2}

\bibitem[noa({[n.\,d.]})]%
        {noauthor_video_nodate}
 \bibinfo{year}{[n.\,d.]}\natexlab{}.
\newblock \bibinfo{title}{Video {Analytics} {Market} {Size}, {Growth}
  {\textbar} {Global} {Report} [2022-2029]}.
\newblock
\newblock
\urldef\tempurl%
\url{https://www.fortunebusinessinsights.com/industry-reports/video-analytics-market-101114}
\showURL{%
\tempurl}


\bibitem[noa(2023)]%
        {noauthor_encodermec_2023}
 \bibinfo{year}{2023}\natexlab{}.
\newblock \bibinfo{title}{encoder/me.c · master · {VideoLAN} / x264 ·
  {GitLab}}.
\newblock
\newblock
\urldef\tempurl%
\url{https://code.videolan.org/videolan/x264/-/blob/master/encoder/me.c}
\showURL{%
\tempurl}


\bibitem[{Anil Chandra, Naidu Matcha}({[n.\,d.]})]%
        {anil_chandra_naidu_matcha_2021_nodate}
\bibfield{author}{\bibinfo{person}{{Anil Chandra, Naidu Matcha}}.}
  \bibinfo{year}{[n.\,d.]}\natexlab{}.
\newblock \bibinfo{title}{A 2021 guide to {Semantic} {Segmentation}}.
\newblock
\newblock
\urldef\tempurl%
\url{https://nanonets.com/blog/semantic-image-segmentation-2020/}
\showURL{%
\tempurl}


\bibitem[Baheti et~al\mbox{.}(2020)]%
        {Baheti_2020_CVPR_Workshops}
\bibfield{author}{\bibinfo{person}{Bhakti Baheti}, \bibinfo{person}{Shubham
  Innani}, \bibinfo{person}{Suhas Gajre}, {and} \bibinfo{person}{Sanjay
  Talbar}.} \bibinfo{year}{2020}\natexlab{}.
\newblock \showarticletitle{Eff-UNet: A Novel Architecture for Semantic
  Segmentation in Unstructured Environment}. In
  \bibinfo{booktitle}{\emph{Proceedings of the IEEE/CVF Conference on Computer
  Vision and Pattern Recognition (CVPR) Workshops}}.
\newblock


\bibitem[Bhatnagar et~al\mbox{.}(2020)]%
        {bhatnagar_drone_2020}
\bibfield{author}{\bibinfo{person}{Saheba Bhatnagar}, \bibinfo{person}{Laurence
  Gill}, {and} \bibinfo{person}{Bidisha Ghosh}.}
  \bibinfo{year}{2020}\natexlab{}.
\newblock \showarticletitle{Drone {Image} {Segmentation} {Using} {Machine} and
  {Deep} {Learning} for {Mapping} {Raised} {Bog} {Vegetation} {Communities}}.
\newblock \bibinfo{journal}{\emph{Remote Sensing}} \bibinfo{volume}{12},
  \bibinfo{number}{16} (\bibinfo{date}{Jan.} \bibinfo{year}{2020}),
  \bibinfo{pages}{2602}.
\newblock
\showISSN{2072-4292}
\urldef\tempurl%
\url{https://doi.org/10.3390/rs12162602}
\showDOI{\tempurl}
\newblock
\shownote{Number: 16 Publisher: Multidisciplinary Digital Publishing
  Institute}.


\bibitem[Calian et~al\mbox{.}(2020)]%
        {calian2020balancing}
\bibfield{author}{\bibinfo{person}{Dan~A Calian}, \bibinfo{person}{Daniel~J
  Mankowitz}, \bibinfo{person}{Tom Zahavy}, \bibinfo{person}{Zhongwen Xu},
  \bibinfo{person}{Junhyuk Oh}, \bibinfo{person}{Nir Levine}, {and}
  \bibinfo{person}{Timothy Mann}.} \bibinfo{year}{2020}\natexlab{}.
\newblock \showarticletitle{Balancing constraints and rewards with
  meta-gradient d4pg}.
\newblock \bibinfo{journal}{\emph{arXiv preprint arXiv:2010.06324}}
  (\bibinfo{year}{2020}).
\newblock


\bibitem[Can et~al\mbox{.}(2021)]%
        {Can_2021_ICCV}
\bibfield{author}{\bibinfo{person}{Yigit~Baran Can}, \bibinfo{person}{Alexander
  Liniger}, \bibinfo{person}{Danda~Pani Paudel}, {and} \bibinfo{person}{Luc
  Van~Gool}.} \bibinfo{year}{2021}\natexlab{}.
\newblock \showarticletitle{Structured Bird's-Eye-View Traffic Scene
  Understanding From Onboard Images}. In \bibinfo{booktitle}{\emph{Proceedings
  of the IEEE/CVF International Conference on Computer Vision (ICCV)}}.
  \bibinfo{pages}{15661--15670}.
\newblock


\bibitem[Canel et~al\mbox{.}(2019)]%
        {canel_scaling_2019}
\bibfield{author}{\bibinfo{person}{Christopher Canel}, \bibinfo{person}{Thomas
  Kim}, \bibinfo{person}{Giulio Zhou}, \bibinfo{person}{Conglong Li},
  \bibinfo{person}{Hyeontaek Lim}, \bibinfo{person}{David~G Andersen},
  \bibinfo{person}{Michael Kaminsky}, {and} \bibinfo{person}{Subramanya~R
  Dulloor}.} \bibinfo{year}{2019}\natexlab{}.
\newblock \showarticletitle{Scaling {Video} {Analytics} on {Constrained} {Edge}
  {Nodes}}. In \bibinfo{booktitle}{\emph{{SysML}}}. \bibinfo{pages}{12}.
\newblock


\bibitem[Chakravarthy et~al\mbox{.}(2022)]%
        {chakravarthy_dronesegnet_2022}
\bibfield{author}{\bibinfo{person}{Anirudh~S. Chakravarthy},
  \bibinfo{person}{Soumendu Sinha}, \bibinfo{person}{Pratik Narang},
  \bibinfo{person}{Murari Mandal}, \bibinfo{person}{Vinay Chamola}, {and}
  \bibinfo{person}{F.~Richard Yu}.} \bibinfo{year}{2022}\natexlab{}.
\newblock \showarticletitle{{DroneSegNet}: {Robust} {Aerial} {Semantic}
  {Segmentation} for {UAV}-{Based} {IoT} {Applications}}.
\newblock \bibinfo{journal}{\emph{IEEE Transactions on Vehicular Technology}}
  \bibinfo{volume}{71}, \bibinfo{number}{4} (\bibinfo{date}{April}
  \bibinfo{year}{2022}), \bibinfo{pages}{4277--4286}.
\newblock
\showISSN{1939-9359}
\urldef\tempurl%
\url{https://doi.org/10.1109/TVT.2022.3144358}
\showDOI{\tempurl}
\newblock
\shownote{Conference Name: IEEE Transactions on Vehicular Technology}.


\bibitem[Chandra et~al\mbox{.}(2018)]%
        {vss-acc-chandra2018deep}
\bibfield{author}{\bibinfo{person}{Siddhartha Chandra},
  \bibinfo{person}{Camille Couprie}, {and} \bibinfo{person}{Iasonas Kokkinos}.}
  \bibinfo{year}{2018}\natexlab{}.
\newblock \showarticletitle{Deep spatio-temporal random fields for efficient
  video segmentation}. In \bibinfo{booktitle}{\emph{Proceedings of the IEEE
  Conference on Computer Vision and Pattern Recognition}}.
  \bibinfo{pages}{8915--8924}.
\newblock


\bibitem[Chen et~al\mbox{.}(2019)]%
        {chen_importance-aware_2019}
\bibfield{author}{\bibinfo{person}{Bike Chen}, \bibinfo{person}{Chen Gong},
  {and} \bibinfo{person}{Jian Yang}.} \bibinfo{year}{2019}\natexlab{}.
\newblock \showarticletitle{Importance-{Aware} {Semantic} {Segmentation} for
  {Autonomous} {Vehicles}}.
\newblock \bibinfo{journal}{\emph{IEEE Transactions on Intelligent
  Transportation Systems}} \bibinfo{volume}{20}, \bibinfo{number}{1}
  (\bibinfo{date}{Jan.} \bibinfo{year}{2019}), \bibinfo{pages}{137--148}.
\newblock
\showISSN{1558-0016}
\urldef\tempurl%
\url{https://doi.org/10.1109/TITS.2018.2801309}
\showDOI{\tempurl}
\newblock
\shownote{Conference Name: IEEE Transactions on Intelligent Transportation
  Systems}.


\bibitem[Chen et~al\mbox{.}(2015)]%
        {chen_glimpse_2015}
\bibfield{author}{\bibinfo{person}{Tiffany Yu-Han Chen}, \bibinfo{person}{Lenin
  Ravindranath}, \bibinfo{person}{Shuo Deng}, \bibinfo{person}{Paramvir Bahl},
  {and} \bibinfo{person}{Hari Balakrishnan}.} \bibinfo{year}{2015}\natexlab{}.
\newblock \showarticletitle{Glimpse: {Continuous}, {Real}-{Time} {Object}
  {Recognition} on {Mobile} {Devices}}. In
  \bibinfo{booktitle}{\emph{Proceedings of the 13th {ACM} {Conference} on
  {Embedded} {Networked} {Sensor} {Systems}}}. \bibinfo{publisher}{ACM},
  \bibinfo{address}{Seoul South Korea}, \bibinfo{pages}{155--168}.
\newblock
\showISBNx{978-1-4503-3631-4}
\urldef\tempurl%
\url{https://doi.org/10.1145/2809695.2809711}
\showDOI{\tempurl}


\bibitem[Ditschuneit and Otterbach(2022)]%
        {ditschuneit_auto-compressing_2022}
\bibfield{author}{\bibinfo{person}{Konstantin Ditschuneit} {and}
  \bibinfo{person}{Johannes~S. Otterbach}.} \bibinfo{year}{2022}\natexlab{}.
\newblock \bibinfo{title}{Auto-{Compressing} {Subset} {Pruning} for {Semantic}
  {Image} {Segmentation}}.
\newblock
\newblock
\urldef\tempurl%
\url{https://doi.org/10.48550/arXiv.2201.11103}
\showDOI{\tempurl}
\newblock
\shownote{arXiv:2201.11103 [cs, stat]}.


\bibitem[Du et~al\mbox{.}(2020)]%
        {du_server-driven_2020}
\bibfield{author}{\bibinfo{person}{Kuntai Du}, \bibinfo{person}{Ahsan Pervaiz},
  \bibinfo{person}{Xin Yuan}, \bibinfo{person}{Aakanksha Chowdhery},
  \bibinfo{person}{Qizheng Zhang}, \bibinfo{person}{Henry Hoffmann}, {and}
  \bibinfo{person}{Junchen Jiang}.} \bibinfo{year}{2020}\natexlab{}.
\newblock \showarticletitle{Server-{Driven} {Video} {Streaming} for {Deep}
  {Learning} {Inference}}. In \bibinfo{booktitle}{\emph{Proceedings of the
  {Annual} conference of the {ACM} {Special} {Interest} {Group} on {Data}
  {Communication} on the applications, technologies, architectures, and
  protocols for computer communication}}. \bibinfo{publisher}{ACM},
  \bibinfo{address}{Virtual Event USA}, \bibinfo{pages}{557--570}.
\newblock
\showISBNx{978-1-4503-7955-7}
\urldef\tempurl%
\url{https://doi.org/10.1145/3387514.3405887}
\showDOI{\tempurl}


\bibitem[Feng et~al\mbox{.}(2021)]%
        {feng_deep_2021}
\bibfield{author}{\bibinfo{person}{Di Feng}, \bibinfo{person}{Christian
  Haase-Schütz}, \bibinfo{person}{Lars Rosenbaum}, \bibinfo{person}{Heinz
  Hertlein}, \bibinfo{person}{Claudius Gläser}, \bibinfo{person}{Fabian Timm},
  \bibinfo{person}{Werner Wiesbeck}, {and} \bibinfo{person}{Klaus Dietmayer}.}
  \bibinfo{year}{2021}\natexlab{}.
\newblock \showarticletitle{Deep {Multi}-{Modal} {Object} {Detection} and
  {Semantic} {Segmentation} for {Autonomous} {Driving}: {Datasets}, {Methods},
  and {Challenges}}.
\newblock \bibinfo{journal}{\emph{IEEE Transactions on Intelligent
  Transportation Systems}} \bibinfo{volume}{22}, \bibinfo{number}{3}
  (\bibinfo{date}{March} \bibinfo{year}{2021}), \bibinfo{pages}{1341--1360}.
\newblock
\showISSN{1558-0016}
\urldef\tempurl%
\url{https://doi.org/10.1109/TITS.2020.2972974}
\showDOI{\tempurl}
\newblock
\shownote{Conference Name: IEEE Transactions on Intelligent Transportation
  Systems}.


\bibitem[Gadde et~al\mbox{.}(2017)]%
        {vss-acc-gadde2017semantic}
\bibfield{author}{\bibinfo{person}{Raghudeep Gadde}, \bibinfo{person}{Varun
  Jampani}, {and} \bibinfo{person}{Peter~V Gehler}.}
  \bibinfo{year}{2017}\natexlab{}.
\newblock \showarticletitle{Semantic video cnns through representation
  warping}. In \bibinfo{booktitle}{\emph{Proceedings of the IEEE International
  Conference on Computer Vision}}. \bibinfo{pages}{4453--4462}.
\newblock


\bibitem[Guerrero~Tello et~al\mbox{.}(2022)]%
        {guerrero_tello_convolutional_2022}
\bibfield{author}{\bibinfo{person}{José~Francisco Guerrero~Tello},
  \bibinfo{person}{Mauro Coltelli}, \bibinfo{person}{Maria Marsella},
  \bibinfo{person}{Angela Celauro}, {and} \bibinfo{person}{José~Antonio
  Palenzuela~Baena}.} \bibinfo{year}{2022}\natexlab{}.
\newblock \showarticletitle{Convolutional {Neural} {Network} {Algorithms} for
  {Semantic} {Segmentation} of {Volcanic} {Ash} {Plumes} {Using} {Visible}
  {Camera} {Imagery}}.
\newblock \bibinfo{journal}{\emph{Remote Sensing}} \bibinfo{volume}{14},
  \bibinfo{number}{18} (\bibinfo{date}{Jan.} \bibinfo{year}{2022}),
  \bibinfo{pages}{4477}.
\newblock
\showISSN{2072-4292}
\urldef\tempurl%
\url{https://doi.org/10.3390/rs14184477}
\showDOI{\tempurl}
\newblock
\shownote{Number: 18 Publisher: Multidisciplinary Digital Publishing
  Institute}.


\bibitem[Guo et~al\mbox{.}(2018)]%
        {guo_foggycache_2018}
\bibfield{author}{\bibinfo{person}{Peizhen Guo}, \bibinfo{person}{Bo Hu},
  \bibinfo{person}{Rui Li}, {and} \bibinfo{person}{Wenjun Hu}.}
  \bibinfo{year}{2018}\natexlab{}.
\newblock \showarticletitle{{FoggyCache}: {Cross}-{Device} {Approximate}
  {Computation} {Reuse}}. In \bibinfo{booktitle}{\emph{Proceedings of the 24th
  {Annual} {International} {Conference} on {Mobile} {Computing} and
  {Networking}}} \emph{(\bibinfo{series}{{MobiCom} '18})}.
  \bibinfo{publisher}{Association for Computing Machinery},
  \bibinfo{address}{New York, NY, USA}, \bibinfo{pages}{19--34}.
\newblock
\showISBNx{978-1-4503-5903-0}
\urldef\tempurl%
\url{https://doi.org/10.1145/3241539.3241557}
\showDOI{\tempurl}


\bibitem[Hendrycks and Gimpel(2018)]%
        {hendrycks_baseline_2018}
\bibfield{author}{\bibinfo{person}{Dan Hendrycks} {and} \bibinfo{person}{Kevin
  Gimpel}.} \bibinfo{year}{2018}\natexlab{}.
\newblock \bibinfo{title}{A {Baseline} for {Detecting} {Misclassified} and
  {Out}-of-{Distribution} {Examples} in {Neural} {Networks}}.
\newblock
\newblock
\urldef\tempurl%
\url{https://doi.org/10.48550/arXiv.1610.02136}
\showDOI{\tempurl}
\newblock
\shownote{arXiv:1610.02136 [cs]}.


\bibitem[Hu et~al\mbox{.}(2020)]%
        {vss-cost-hu2020temporally}
\bibfield{author}{\bibinfo{person}{Ping Hu}, \bibinfo{person}{Fabian Caba},
  \bibinfo{person}{Oliver Wang}, \bibinfo{person}{Zhe Lin},
  \bibinfo{person}{Stan Sclaroff}, {and} \bibinfo{person}{Federico Perazzi}.}
  \bibinfo{year}{2020}\natexlab{}.
\newblock \showarticletitle{Temporally distributed networks for fast video
  semantic segmentation}. In \bibinfo{booktitle}{\emph{Proceedings of the
  IEEE/CVF Conference on Computer Vision and Pattern Recognition}}.
  \bibinfo{pages}{8818--8827}.
\newblock


\bibitem[Jiang et~al\mbox{.}(2018)]%
        {jiang_chameleon_2018}
\bibfield{author}{\bibinfo{person}{Junchen Jiang}, \bibinfo{person}{Ganesh
  Ananthanarayanan}, \bibinfo{person}{Peter Bodik}, \bibinfo{person}{Siddhartha
  Sen}, {and} \bibinfo{person}{Ion Stoica}.} \bibinfo{year}{2018}\natexlab{}.
\newblock \showarticletitle{Chameleon: scalable adaptation of video analytics}.
  In \bibinfo{booktitle}{\emph{{SIGCOM}}} \emph{(\bibinfo{series}{{SIGCOMM}
  '18})}. \bibinfo{publisher}{Association for Computing Machinery},
  \bibinfo{address}{New York, NY, USA}, \bibinfo{pages}{253--266}.
\newblock
\showISBNx{978-1-4503-5567-4}
\urldef\tempurl%
\url{https://doi.org/10.1145/3230543.3230574}
\showDOI{\tempurl}


\bibitem[Jiang et~al\mbox{.}(2021)]%
        {jiang_joint_2021}
\bibfield{author}{\bibinfo{person}{Jingyan Jiang}, \bibinfo{person}{Ziyue Luo},
  \bibinfo{person}{Chenghao Hu}, \bibinfo{person}{Zhaoliang He},
  \bibinfo{person}{Zhi Wang}, \bibinfo{person}{Shutao Xia}, {and}
  \bibinfo{person}{Chuan Wu}.} \bibinfo{year}{2021}\natexlab{}.
\newblock \showarticletitle{Joint {Model} and {Data} {Adaptation} for {Cloud}
  {Inference} {Serving}}. In \bibinfo{booktitle}{\emph{2021 {IEEE}
  {Real}-{Time} {Systems} {Symposium} ({RTSS})}}. \bibinfo{pages}{279--289}.
\newblock
\urldef\tempurl%
\url{https://doi.org/10.1109/RTSS52674.2021.00034}
\showDOI{\tempurl}
\newblock
\shownote{ISSN: 2576-3172}.


\bibitem[Jin et~al\mbox{.}(2017)]%
        {vss-acc-jin2017video}
\bibfield{author}{\bibinfo{person}{Xiaojie Jin}, \bibinfo{person}{Xin Li},
  \bibinfo{person}{Huaxin Xiao}, \bibinfo{person}{Xiaohui Shen},
  \bibinfo{person}{Zhe Lin}, \bibinfo{person}{Jimei Yang},
  \bibinfo{person}{Yunpeng Chen}, \bibinfo{person}{Jian Dong},
  \bibinfo{person}{Luoqi Liu}, \bibinfo{person}{Zequn Jie}, {et~al\mbox{.}}}
  \bibinfo{year}{2017}\natexlab{}.
\newblock \showarticletitle{Video scene parsing with predictive feature
  learning}. In \bibinfo{booktitle}{\emph{Proceedings of the IEEE International
  Conference on Computer Vision}}. \bibinfo{pages}{5580--5588}.
\newblock


\bibitem[Kang et~al\mbox{.}(2017)]%
        {kang_noscope_2017}
\bibfield{author}{\bibinfo{person}{Daniel Kang}, \bibinfo{person}{John Emmons},
  \bibinfo{person}{Firas Abuzaid}, \bibinfo{person}{Peter Bailis}, {and}
  \bibinfo{person}{Matei Zaharia}.} \bibinfo{year}{2017}\natexlab{}.
\newblock \showarticletitle{{NoScope}: optimizing neural network queries over
  video at scale}.
\newblock \bibinfo{journal}{\emph{Proceedings of the VLDB Endowment}}
  \bibinfo{volume}{10}, \bibinfo{number}{11} (\bibinfo{date}{Aug.}
  \bibinfo{year}{2017}), \bibinfo{pages}{1586--1597}.
\newblock
\showISSN{2150-8097}
\urldef\tempurl%
\url{https://doi.org/10.14778/3137628.3137664}
\showDOI{\tempurl}


\bibitem[Li et~al\mbox{.}(2020)]%
        {li_reducto_2020}
\bibfield{author}{\bibinfo{person}{Yuanqi Li}, \bibinfo{person}{Arthi
  Padmanabhan}, \bibinfo{person}{Pengzhan Zhao}, \bibinfo{person}{Yufei Wang},
  \bibinfo{person}{Guoqing~Harry Xu}, {and} \bibinfo{person}{Ravi Netravali}.}
  \bibinfo{year}{2020}\natexlab{}.
\newblock \showarticletitle{Reducto: {On}-{Camera} {Filtering} for
  {Resource}-{Efficient} {Real}-{Time} {Video} {Analytics}}. In
  \bibinfo{booktitle}{\emph{Proceedings of the {Annual} conference of the {ACM}
  {Special} {Interest} {Group} on {Data} {Communication} on the applications,
  technologies, architectures, and protocols for computer communication}}
  \emph{(\bibinfo{series}{{SIGCOMM} '20})}. \bibinfo{publisher}{Association for
  Computing Machinery}, \bibinfo{address}{New York, NY, USA},
  \bibinfo{pages}{359--376}.
\newblock
\showISBNx{978-1-4503-7955-7}
\urldef\tempurl%
\url{https://doi.org/10.1145/3387514.3405874}
\showDOI{\tempurl}


\bibitem[Li et~al\mbox{.}(2018)]%
        {vss-cost-li2018low}
\bibfield{author}{\bibinfo{person}{Yule Li}, \bibinfo{person}{Jianping Shi},
  {and} \bibinfo{person}{Dahua Lin}.} \bibinfo{year}{2018}\natexlab{}.
\newblock \showarticletitle{Low-latency video semantic segmentation}. In
  \bibinfo{booktitle}{\emph{Proceedings of the IEEE Conference on Computer
  Vision and Pattern Recognition}}. \bibinfo{pages}{5997--6005}.
\newblock


\bibitem[Liu et~al\mbox{.}(2022)]%
        {liu_adamask_2022}
\bibfield{author}{\bibinfo{person}{Shengzhong Liu}, \bibinfo{person}{Tianshi
  Wang}, \bibinfo{person}{Jinyang Li}, \bibinfo{person}{Dachun Sun},
  \bibinfo{person}{Mani Srivastava}, {and} \bibinfo{person}{Tarek Abdelzaher}.}
  \bibinfo{year}{2022}\natexlab{}.
\newblock \showarticletitle{{AdaMask}: {Enabling} {Machine}-{Centric} {Video}
  {Streaming} with {Adaptive} {Frame} {Masking} for {DNN} {Inference}
  {Offloading}}. In \bibinfo{booktitle}{\emph{Proceedings of the 30th {ACM}
  {International} {Conference} on {Multimedia}}} \emph{(\bibinfo{series}{{MM}
  '22})}. \bibinfo{publisher}{Association for Computing Machinery},
  \bibinfo{address}{New York, NY, USA}, \bibinfo{pages}{3035--3044}.
\newblock
\showISBNx{978-1-4503-9203-7}
\urldef\tempurl%
\url{https://doi.org/10.1145/3503161.3548033}
\showDOI{\tempurl}


\bibitem[Ltd({[n.\,d.]})]%
        {ltd_raspberry_nodate}
\bibfield{author}{\bibinfo{person}{Raspberry~Pi Ltd}.}
  \bibinfo{year}{[n.\,d.]}\natexlab{}.
\newblock \bibinfo{title}{Raspberry {Pi} 4 {Model} {B} specifications}.
\newblock
\newblock
\urldef\tempurl%
\url{https://www.raspberrypi.com/products/raspberry-pi-4-model-b/specifications/}
\showURL{%
\tempurl}


\bibitem[Mahasseni et~al\mbox{.}(2017)]%
        {vss-cost-mahasseni2017budget}
\bibfield{author}{\bibinfo{person}{Behrooz Mahasseni}, \bibinfo{person}{Sinisa
  Todorovic}, {and} \bibinfo{person}{Alan Fern}.}
  \bibinfo{year}{2017}\natexlab{}.
\newblock \showarticletitle{Budget-aware deep semantic video segmentation}. In
  \bibinfo{booktitle}{\emph{Proceedings of the IEEE Conference on Computer
  Vision and Pattern Recognition}}. \bibinfo{pages}{1029--1038}.
\newblock


\bibitem[Mandhane et~al\mbox{.}(2022)]%
        {mandhane_muzero_2022}
\bibfield{author}{\bibinfo{person}{Amol Mandhane}, \bibinfo{person}{Anton
  Zhernov}, \bibinfo{person}{Maribeth Rauh}, \bibinfo{person}{Chenjie Gu},
  \bibinfo{person}{Miaosen Wang}, \bibinfo{person}{Flora Xue},
  \bibinfo{person}{Wendy Shang}, \bibinfo{person}{Derek Pang},
  \bibinfo{person}{Rene Claus}, \bibinfo{person}{Ching-Han Chiang},
  \bibinfo{person}{Cheng Chen}, \bibinfo{person}{Jingning Han},
  \bibinfo{person}{Angie Chen}, \bibinfo{person}{Daniel~J. Mankowitz},
  \bibinfo{person}{Jackson Broshear}, \bibinfo{person}{Julian Schrittwieser},
  \bibinfo{person}{Thomas Hubert}, \bibinfo{person}{Oriol Vinyals}, {and}
  \bibinfo{person}{Timothy Mann}.} \bibinfo{year}{2022}\natexlab{}.
\newblock \bibinfo{title}{{MuZero} with {Self}-competition for {Rate} {Control}
  in {VP9} {Video} {Compression}}.
\newblock
\newblock
\urldef\tempurl%
\url{https://doi.org/10.48550/arXiv.2202.06626}
\showDOI{\tempurl}
\newblock
\shownote{arXiv:2202.06626 [cs, eess]}.


\bibitem[Mnih et~al\mbox{.}(2016)]%
        {mnih2016asynchronous}
\bibfield{author}{\bibinfo{person}{Volodymyr Mnih},
  \bibinfo{person}{Adria~Puigdomenech Badia}, \bibinfo{person}{Mehdi Mirza},
  \bibinfo{person}{Alex Graves}, \bibinfo{person}{Timothy Lillicrap},
  \bibinfo{person}{Tim Harley}, \bibinfo{person}{David Silver}, {and}
  \bibinfo{person}{Koray Kavukcuoglu}.} \bibinfo{year}{2016}\natexlab{}.
\newblock \showarticletitle{Asynchronous methods for deep reinforcement
  learning}. In \bibinfo{booktitle}{\emph{International conference on machine
  learning}}. PMLR, \bibinfo{pages}{1928--1937}.
\newblock


\bibitem[Mo et~al\mbox{.}(2022)]%
        {mo_review_2022}
\bibfield{author}{\bibinfo{person}{Yujian Mo}, \bibinfo{person}{Yan Wu},
  \bibinfo{person}{Xinneng Yang}, \bibinfo{person}{Feilin Liu}, {and}
  \bibinfo{person}{Yujun Liao}.} \bibinfo{year}{2022}\natexlab{}.
\newblock \showarticletitle{Review the state-of-the-art technologies of
  semantic segmentation based on deep learning}.
\newblock \bibinfo{journal}{\emph{Neurocomputing}}  \bibinfo{volume}{493}
  (\bibinfo{date}{July} \bibinfo{year}{2022}), \bibinfo{pages}{626--646}.
\newblock
\showISSN{09252312}
\urldef\tempurl%
\url{https://doi.org/10.1016/j.neucom.2022.01.005}
\showDOI{\tempurl}


\bibitem[Muhadi et~al\mbox{.}(2021)]%
        {muhadi2021deep}
\bibfield{author}{\bibinfo{person}{Nur~Atirah Muhadi},
  \bibinfo{person}{Ahmad~Fikri Abdullah}, \bibinfo{person}{Siti~Khairunniza
  Bejo}, \bibinfo{person}{Muhammad~Razif Mahadi}, {and} \bibinfo{person}{Ana
  Mijic}.} \bibinfo{year}{2021}\natexlab{}.
\newblock \showarticletitle{Deep learning semantic segmentation for water level
  estimation using surveillance camera}.
\newblock \bibinfo{journal}{\emph{Applied Sciences}} \bibinfo{volume}{11},
  \bibinfo{number}{20} (\bibinfo{year}{2021}), \bibinfo{pages}{9691}.
\newblock


\bibitem[murad et~al\mbox{.}(2022)]%
        {murad_dao_2022}
\bibfield{author}{\bibinfo{person}{taslim murad}, \bibinfo{person}{Anh Nguyen},
  {and} \bibinfo{person}{Zhisheng Yan}.} \bibinfo{year}{2022}\natexlab{}.
\newblock \showarticletitle{{DAO}: {Dynamic} {Adaptive} {Offloading} for
  {Video} {Analytics}}. In \bibinfo{booktitle}{\emph{Proceedings of the 30th
  {ACM} {International} {Conference} on {Multimedia}}}
  \emph{(\bibinfo{series}{{MM} '22})}. \bibinfo{publisher}{Association for
  Computing Machinery}, \bibinfo{address}{New York, NY, USA},
  \bibinfo{pages}{3017--3025}.
\newblock
\showISBNx{978-1-4503-9203-7}
\urldef\tempurl%
\url{https://doi.org/10.1145/3503161.3548249}
\showDOI{\tempurl}


\bibitem[Nilsson and Sminchisescu(2018)]%
        {vss-acc-nilsson2018semantic}
\bibfield{author}{\bibinfo{person}{David Nilsson} {and}
  \bibinfo{person}{Cristian Sminchisescu}.} \bibinfo{year}{2018}\natexlab{}.
\newblock \showarticletitle{Semantic video segmentation by gated recurrent flow
  propagation}. In \bibinfo{booktitle}{\emph{Proceedings of the IEEE conference
  on computer vision and pattern recognition}}. \bibinfo{pages}{6819--6828}.
\newblock


\bibitem[Paternain et~al\mbox{.}(2019)]%
        {paternain2019constrained}
\bibfield{author}{\bibinfo{person}{Santiago Paternain}, \bibinfo{person}{Luiz
  Chamon}, \bibinfo{person}{Miguel Calvo-Fullana}, {and}
  \bibinfo{person}{Alejandro Ribeiro}.} \bibinfo{year}{2019}\natexlab{}.
\newblock \showarticletitle{Constrained reinforcement learning has zero duality
  gap}.
\newblock \bibinfo{journal}{\emph{Advances in Neural Information Processing
  Systems}}  \bibinfo{volume}{32} (\bibinfo{year}{2019}).
\newblock


\bibitem[Piérard et~al\mbox{.}(2023)]%
        {pierard_mixture_2023}
\bibfield{author}{\bibinfo{person}{Sébastien Piérard},
  \bibinfo{person}{Anthony Cioppa}, \bibinfo{person}{Anaïs Halin},
  \bibinfo{person}{Renaud Vandeghen}, \bibinfo{person}{Maxime Zanella},
  \bibinfo{person}{Benoît Macq}, \bibinfo{person}{Saïd Mahmoudi}, {and}
  \bibinfo{person}{Marc Van~Droogenbroeck}.} \bibinfo{year}{2023}\natexlab{}.
\newblock \showarticletitle{Mixture {Domain} {Adaptation} {To} {Improve}
  {Semantic} {Segmentation} in {Real}-{World} {Surveillance}}.
  \bibinfo{pages}{22--31}.
\newblock
\urldef\tempurl%
\url{https://openaccess.thecvf.com/content/WACV2023W/RWS/html/Pierard_Mixture_Domain_Adaptation_To_Improve_Semantic_Segmentation_in_Real-World_Surveillance_WACVW_2023_paper.html}
\showURL{%
\tempurl}


\bibitem[Redmon and Farhadi(2018)]%
        {redmon2018yolov3}
\bibfield{author}{\bibinfo{person}{Joseph Redmon} {and} \bibinfo{person}{Ali
  Farhadi}.} \bibinfo{year}{2018}\natexlab{}.
\newblock \showarticletitle{Yolov3: An incremental improvement}.
\newblock \bibinfo{journal}{\emph{arXiv preprint arXiv:1804.02767}}
  (\bibinfo{year}{2018}).
\newblock


\bibitem[Sandler et~al\mbox{.}(2018)]%
        {sandler2018mobilenetv2}
\bibfield{author}{\bibinfo{person}{Mark Sandler}, \bibinfo{person}{Andrew
  Howard}, \bibinfo{person}{Menglong Zhu}, \bibinfo{person}{Andrey Zhmoginov},
  {and} \bibinfo{person}{Liang-Chieh Chen}.} \bibinfo{year}{2018}\natexlab{}.
\newblock \showarticletitle{Mobilenetv2: Inverted residuals and linear
  bottlenecks}. In \bibinfo{booktitle}{\emph{Proceedings of the IEEE conference
  on computer vision and pattern recognition}}. \bibinfo{pages}{4510--4520}.
\newblock


\bibitem[Schulman et~al\mbox{.}(2017)]%
        {schulman_proximal_2017}
\bibfield{author}{\bibinfo{person}{John Schulman}, \bibinfo{person}{Filip
  Wolski}, \bibinfo{person}{Prafulla Dhariwal}, \bibinfo{person}{Alec Radford},
  {and} \bibinfo{person}{Oleg Klimov}.} \bibinfo{year}{2017}\natexlab{}.
\newblock \showarticletitle{Proximal {Policy} {Optimization} {Algorithms}}. In
  \bibinfo{booktitle}{\emph{{arXiv}:1707.06347 [cs]}}.
\newblock
\urldef\tempurl%
\url{http://arxiv.org/abs/1707.06347}
\showURL{%
\tempurl}
\newblock
\shownote{arXiv: 1707.06347}.


\bibitem[Sutton and Barto(2018)]%
        {sutton2018reinforcement}
\bibfield{author}{\bibinfo{person}{Richard~S Sutton} {and}
  \bibinfo{person}{Andrew~G Barto}.} \bibinfo{year}{2018}\natexlab{}.
\newblock \bibinfo{booktitle}{\emph{Reinforcement learning: An introduction}}.
\newblock \bibinfo{publisher}{MIT press}.
\newblock


\bibitem[Tomar(2006)]%
        {tomar2006converting}
\bibfield{author}{\bibinfo{person}{Suramya Tomar}.}
  \bibinfo{year}{2006}\natexlab{}.
\newblock \showarticletitle{Converting video formats with FFmpeg}.
\newblock \bibinfo{journal}{\emph{Linux Journal}} \bibinfo{volume}{2006},
  \bibinfo{number}{146} (\bibinfo{year}{2006}), \bibinfo{pages}{10}.
\newblock


\bibitem[Wang et~al\mbox{.}(2020)]%
        {wang_joint_2020}
\bibfield{author}{\bibinfo{person}{Can Wang}, \bibinfo{person}{Sheng Zhang},
  \bibinfo{person}{Yu Chen}, \bibinfo{person}{Zhuzhong Qian},
  \bibinfo{person}{Jie Wu}, {and} \bibinfo{person}{Mingjun Xiao}.}
  \bibinfo{year}{2020}\natexlab{}.
\newblock \showarticletitle{Joint {Configuration} {Adaptation} and {Bandwidth}
  {Allocation} for {Edge}-based {Real}-time {Video} {Analytics}}. In
  \bibinfo{booktitle}{\emph{{IEEE} {INFOCOM} 2020 - {IEEE} {Conference} on
  {Computer} {Communications}}}. \bibinfo{pages}{257--266}.
\newblock
\urldef\tempurl%
\url{https://doi.org/10.1109/INFOCOM41043.2020.9155524}
\showDOI{\tempurl}
\newblock
\shownote{ISSN: 2641-9874}.


\bibitem[Wang et~al\mbox{.}(2021)]%
        {wang_swiftnet_2021}
\bibfield{author}{\bibinfo{person}{Haochen Wang}, \bibinfo{person}{Xiaolong
  Jiang}, \bibinfo{person}{Haibing Ren}, \bibinfo{person}{Yao Hu}, {and}
  \bibinfo{person}{Song Bai}.} \bibinfo{year}{2021}\natexlab{}.
\newblock \showarticletitle{{SwiftNet}: {Real}-time {Video} {Object}
  {Segmentation}}. In \bibinfo{booktitle}{\emph{2021 {IEEE}/{CVF} {Conference}
  on {Computer} {Vision} and {Pattern} {Recognition} ({CVPR})}}.
  \bibinfo{publisher}{IEEE}, \bibinfo{address}{Nashville, TN, USA},
  \bibinfo{pages}{1296--1305}.
\newblock
\showISBNx{978-1-66544-509-2}
\urldef\tempurl%
\url{https://doi.org/10.1109/CVPR46437.2021.00135}
\showDOI{\tempurl}


\bibitem[Xiao et~al\mbox{.}(2022)]%
        {xiao_dnn-driven_2022}
\bibfield{author}{\bibinfo{person}{Xuedou Xiao}, \bibinfo{person}{Juecheng
  Zhang}, \bibinfo{person}{Wei Wang}, \bibinfo{person}{Jianhua He}, {and}
  \bibinfo{person}{Qian Zhang}.} \bibinfo{year}{2022}\natexlab{}.
\newblock \showarticletitle{{DNN}-{Driven} {Compressive} {Ofﬂoading} for
  {Edge}-{Assisted} {Semantic} {Video} {Segmentation}}. In
  \bibinfo{booktitle}{\emph{{IEEE} {INFOCOM} 2022 - {IEEE} {Conference} on
  {Computer} {Communications}}}. \bibinfo{pages}{10}.
\newblock


\bibitem[Yu et~al\mbox{.}(2018)]%
        {yu_bisenet_2018}
\bibfield{author}{\bibinfo{person}{Changqian Yu}, \bibinfo{person}{Jingbo
  Wang}, \bibinfo{person}{Chao Peng}, \bibinfo{person}{Changxin Gao},
  \bibinfo{person}{Gang Yu}, {and} \bibinfo{person}{Nong Sang}.}
  \bibinfo{year}{2018}\natexlab{}.
\newblock \showarticletitle{{BiSeNet}: {Bilateral} {Segmentation} {Network} for
  {Real}-time {Semantic} {Segmentation}}. \bibinfo{pages}{325--341}.
\newblock
\urldef\tempurl%
\url{https://openaccess.thecvf.com/content_ECCV_2018/html/Changqian_Yu_BiSeNet_Bilateral_Segmentation_ECCV_2018_paper.html}
\showURL{%
\tempurl}


\bibitem[Yu et~al\mbox{.}(2020)]%
        {bdd100k}
\bibfield{author}{\bibinfo{person}{Fisher Yu}, \bibinfo{person}{Haofeng Chen},
  \bibinfo{person}{Xin Wang}, \bibinfo{person}{Wenqi Xian},
  \bibinfo{person}{Yingying Chen}, \bibinfo{person}{Fangchen Liu},
  \bibinfo{person}{Vashisht Madhavan}, {and} \bibinfo{person}{Trevor Darrell}.}
  \bibinfo{year}{2020}\natexlab{}.
\newblock \showarticletitle{BDD100K: A Diverse Driving Dataset for
  Heterogeneous Multitask Learning}. In \bibinfo{booktitle}{\emph{The IEEE
  Conference on Computer Vision and Pattern Recognition (CVPR)}}.
\newblock


\bibitem[Yuan et~al\mbox{.}(2022)]%
        {yuan_infi_2022}
\bibfield{author}{\bibinfo{person}{Mu Yuan}, \bibinfo{person}{Lan Zhang},
  \bibinfo{person}{Fengxiang He}, \bibinfo{person}{Xueting Tong}, {and}
  \bibinfo{person}{Xiang-Yang Li}.} \bibinfo{year}{2022}\natexlab{}.
\newblock \showarticletitle{{InFi}: end-to-end learnable input filter for
  resource-efficient mobile-centric inference}. In
  \bibinfo{booktitle}{\emph{Proceedings of the 28th {Annual} {International}
  {Conference} on {Mobile} {Computing} {And} {Networking}}}
  \emph{(\bibinfo{series}{{MobiCom} '22})}. \bibinfo{publisher}{Association for
  Computing Machinery}, \bibinfo{address}{New York, NY, USA},
  \bibinfo{pages}{228--241}.
\newblock
\showISBNx{978-1-4503-9181-8}
\urldef\tempurl%
\url{https://doi.org/10.1145/3495243.3517016}
\showDOI{\tempurl}


\bibitem[Zhang et~al\mbox{.}(2018)]%
        {zhang_awstream_2018}
\bibfield{author}{\bibinfo{person}{Ben Zhang}, \bibinfo{person}{Xin Jin},
  \bibinfo{person}{Sylvia Ratnasamy}, \bibinfo{person}{John Wawrzynek}, {and}
  \bibinfo{person}{Edward~A. Lee}.} \bibinfo{year}{2018}\natexlab{}.
\newblock \showarticletitle{{AWStream}: adaptive wide-area streaming
  analytics}. In \bibinfo{booktitle}{\emph{Proceedings of the 2018 {Conference}
  of the {ACM} {Special} {Interest} {Group} on {Data} {Communication}}}
  \emph{(\bibinfo{series}{{SIGCOMM} '18})}. \bibinfo{publisher}{Association for
  Computing Machinery}, \bibinfo{address}{New York, NY, USA},
  \bibinfo{pages}{236--252}.
\newblock
\showISBNx{978-1-4503-5567-4}
\urldef\tempurl%
\url{https://doi.org/10.1145/3230543.3230554}
\showDOI{\tempurl}


\bibitem[Zhang(2023)]%
        {zhang_fast_2023}
\bibfield{author}{\bibinfo{person}{Eric Zhang}.}
  \bibinfo{year}{2023}\natexlab{}.
\newblock \bibinfo{title}{Fast {Semantic} {Segmentation}}.
\newblock
\newblock
\urldef\tempurl%
\url{https://github.com/ekzhang/fastseg}
\showURL{%
\tempurl}
\newblock
\shownote{original-date: 2020-07-22T22:11:27Z}.


\bibitem[Zhang et~al\mbox{.}(2022a)]%
        {zhang_edge_2022}
\bibfield{author}{\bibinfo{person}{Jialin Zhang}, \bibinfo{person}{Xiang
  Huang}, \bibinfo{person}{Jingao Xu}, \bibinfo{person}{Yue Wu},
  \bibinfo{person}{Qiang Ma}, \bibinfo{person}{Xin Miao}, \bibinfo{person}{Li
  Zhang}, \bibinfo{person}{Pengpeng Chen}, {and} \bibinfo{person}{Zheng Yang}.}
  \bibinfo{year}{2022}\natexlab{a}.
\newblock \showarticletitle{Edge {Assisted} {Real}-time {Instance}
  {Segmentation} on {Mobile} {Devices}}. In \bibinfo{booktitle}{\emph{2022
  {IEEE} 42nd {International} {Conference} on {Distributed} {Computing}
  {Systems} ({ICDCS})}}. \bibinfo{pages}{537--547}.
\newblock
\urldef\tempurl%
\url{https://doi.org/10.1109/ICDCS54860.2022.00058}
\showDOI{\tempurl}
\newblock
\shownote{ISSN: 2575-8411}.


\bibitem[Zhang et~al\mbox{.}(2022b)]%
        {zhang_casva_2022}
\bibfield{author}{\bibinfo{person}{Miao Zhang}, \bibinfo{person}{Fangxin Wang},
  {and} \bibinfo{person}{Jiangchuan Liu}.} \bibinfo{year}{2022}\natexlab{b}.
\newblock \showarticletitle{{CASVA}: {Configuration}-{Adaptive} {Streaming} for
  {Live} {Video} {Analytics}}. In \bibinfo{booktitle}{\emph{{IEEE} {INFOCOM}
  2022 - {IEEE} {Conference} on {Computer} {Communications}}}.
  \bibinfo{publisher}{IEEE}, \bibinfo{address}{London, United Kingdom},
  \bibinfo{pages}{2168--2177}.
\newblock
\showISBNx{978-1-66545-822-1}
\urldef\tempurl%
\url{https://doi.org/10.1109/INFOCOM48880.2022.9796875}
\showDOI{\tempurl}


\bibitem[Zhang et~al\mbox{.}(2022c)]%
        {zhang_adaptive_2022}
\bibfield{author}{\bibinfo{person}{Sheng Zhang}, \bibinfo{person}{Can Wang},
  \bibinfo{person}{Yibo Jin}, \bibinfo{person}{Jie Wu},
  \bibinfo{person}{Zhuzhong Qian}, \bibinfo{person}{Mingjun Xiao}, {and}
  \bibinfo{person}{Sanglu Lu}.} \bibinfo{year}{2022}\natexlab{c}.
\newblock \showarticletitle{Adaptive {Configuration} {Selection} and
  {Bandwidth} {Allocation} for {Edge}-{Based} {Video} {Analytics}}.
\newblock \bibinfo{journal}{\emph{IEEE/ACM Transactions on Networking}}
  \bibinfo{volume}{30}, \bibinfo{number}{1} (\bibinfo{date}{Feb.}
  \bibinfo{year}{2022}), \bibinfo{pages}{285--298}.
\newblock
\showISSN{1558-2566}
\urldef\tempurl%
\url{https://doi.org/10.1109/TNET.2021.3106937}
\showDOI{\tempurl}
\newblock
\shownote{Conference Name: IEEE/ACM Transactions on Networking}.


\bibitem[Zhao et~al\mbox{.}(2018)]%
        {zhao_icnet_2018}
\bibfield{author}{\bibinfo{person}{Hengshuang Zhao}, \bibinfo{person}{Xiaojuan
  Qi}, \bibinfo{person}{Xiaoyong Shen}, \bibinfo{person}{Jianping Shi}, {and}
  \bibinfo{person}{Jiaya Jia}.} \bibinfo{year}{2018}\natexlab{}.
\newblock \bibinfo{title}{{ICNet} for {Real}-{Time} {Semantic} {Segmentation}
  on {High}-{Resolution} {Images}}.
\newblock
\newblock
\urldef\tempurl%
\url{https://doi.org/10.48550/arXiv.1704.08545}
\showDOI{\tempurl}
\newblock
\shownote{arXiv:1704.08545 [cs]}.


\bibitem[Zhao et~al\mbox{.}(2017)]%
        {zhao_pyramid_2017}
\bibfield{author}{\bibinfo{person}{Hengshuang Zhao}, \bibinfo{person}{Jianping
  Shi}, \bibinfo{person}{Xiaojuan Qi}, \bibinfo{person}{Xiaogang Wang}, {and}
  \bibinfo{person}{Jiaya Jia}.} \bibinfo{year}{2017}\natexlab{}.
\newblock \bibinfo{title}{Pyramid {Scene} {Parsing} {Network}}.
\newblock
\newblock
\urldef\tempurl%
\url{https://doi.org/10.48550/arXiv.1612.01105}
\showDOI{\tempurl}
\newblock
\shownote{arXiv:1612.01105 [cs]}.


\bibitem[Zhu et~al\mbox{.}(2017)]%
        {vss-cost-zhu2017deep}
\bibfield{author}{\bibinfo{person}{Xizhou Zhu}, \bibinfo{person}{Yuwen Xiong},
  \bibinfo{person}{Jifeng Dai}, \bibinfo{person}{Lu Yuan}, {and}
  \bibinfo{person}{Yichen Wei}.} \bibinfo{year}{2017}\natexlab{}.
\newblock \showarticletitle{Deep feature flow for video recognition}. In
  \bibinfo{booktitle}{\emph{Proceedings of the IEEE conference on computer
  vision and pattern recognition}}. \bibinfo{pages}{2349--2358}.
\newblock


\end{thebibliography}

\end{document}